\documentclass[twocolumn]{aastex63}

\shortauthors{Duan \& Guo 2019}

\begin{document}
\shorttitle{The Energy Coupling Efficiency of AGN Outbursts}
\title{On the Energy Coupling Efficiency  of AGN Outbursts in Galaxy Clusters}

\correspondingauthor{Fulai Guo}
\email{fulai@shao.ac.cn}

\author{Xiaodong Duan}
\affiliation{Key Laboratory for Research in Galaxies and Cosmology, Shanghai Astronomical Observatory, Chinese Academy of Sciences\\
80 Nandan Road, Shanghai 200030, China}
\affiliation{School of Astronomy and Space Science, University of Chinese Academy of Sciences \\
19A Yuquan Road, Beijing 100049, China}

\author{Fulai Guo}
\affiliation{Key Laboratory for Research in Galaxies and Cosmology, Shanghai Astronomical Observatory, Chinese Academy of Sciences\\
80 Nandan Road, Shanghai 200030, China}
\affiliation{School of Astronomy and Space Science, University of Chinese Academy of Sciences \\
19A Yuquan Road, Beijing 100049, China}

\begin{abstract}

Active galactic nucleus (AGN) jets are believed to be important in solving the cooling flow problem in the intracluster medium (ICM), while the detailed mechanism is still in debate. Here we present a systematic study on the energy coupling efficiency $\eta_{\rm cp}$, the fraction of AGN jet energy transferred to the ICM. We first estimate the values of $\eta_{\rm cp}$ analytically in two extreme cases, which are further confirmed and extended with a parameter study of spherical outbursts in a uniform medium using hydrodynamic simulations. We find that $\eta_{\rm cp}$ increases from $\sim 0.4$ for a weak isobaric injection to $\gtrsim 0.8$ for a powerful point injection. For any given outburst energy, we find two characteristic outburst powers that separate these two extreme cases. We then investigate the energy coupling efficiency of AGN jet outbursts in a realistic ICM with hydrodynamic simulations, finding that jet outbursts are intrinsically different from spherical outbursts. For both powerful and weak jet outbursts, $\eta_{\rm cp}$ is typically around $0.7-0.9$, partly due to the non-spherical nature of jet outbursts, which produce backflows emanating from the hotspots, significantly enhancing the ejecta-ICM interaction. While for powerful outbursts a dominant fraction of the energy transferred from the jet to the ICM is dissipated by shocks, shock dissipation only accounts for $\lesssim 30\%$ of the injected jet energy for weak outbursts. While both powerful and weak outbursts could efficiently heat cooling flows, powerful thermal-energy-dominated jets are most effective in delaying the onset of the central cooling catastrophe.                     

\end{abstract}

\keywords{
 galaxies: active -- galaxies: jets -- galaxies: clusters: intracluster medium -- hydrodynamics -- methods: numerical -- X-rays: galaxies: clusters
 }

\section{Introduction}
\label{section1}
The timescale of radiative cooling in the intracluster medium (ICM), especially in cool cores of galaxy clusters, is much shorter than the typical age of galaxy clusters, and without heating sources, this can result in mass inflow rates and star formation rates much higher than observed, which is referred as the cooling flow problem (\citealt{fabian94}, \citealt{peterson06}; \citealt{mcnamara16}; \citealt{hogan17}; \citealt{lakhchaura18}). Observations of radio jets, X-ray cavities (\citealt{boehringer93}; \citealt{fabian02}; \citealt{birzan04}; \citealt{croston11}, \citealt{vagshette19}), shock structures (\citealt{randall11}; \citealt{randall15}; \citealt{fabian06}; \citealt{vagshette19}) and sound waves (\citealt{fabian03}; \citealt{fabian17}) in many galaxy groups and clusters suggest that the interaction between active galactic nucleus (AGN) jets and the ICM may play an important role in heating the ICM and solving the cooling flow problem (\citealt{owen00}; \citealt{birzan04}; \citealt{mcnamara05}; \citealt{rafferty06}; \citealt{mcnamara07}; \citealt{guo08b}; \citealt{fabian12};  \citealt{mcnamara12}; \citealt{heckman14}; \citealt{li17}; \citealt{martizzi19}). While this AGN jet feedback scenario is widely accepted (\citealt{mcnamara12}; \citealt{soker16}; \citealt{werner19}), the detailed mechanism through which the jet energy is transported to the entire cool cluster core and subsequently dissipated there is still highly debated. The proposed mechanisms include shock heating (\citealt{bruggen07}; \citealt{randall15}; \citealt{li17}; \citealt{guo18}), sound wave dissipation (\citealt{ruszkowski04}; \citealt{fabian17}; \citealt{bambic19}), mixing of the jet ejecta with the ICM (\citealt{hillel16}; \citealt{hillel17}), turbulence dissipation (\citealt{fujita05}; \citealt{enblin06}; \citealt{kunz11}; \citealt{zhuravleva14}; \citealt{zhuravleva16}), and cosmic ray heating (\citealt{guo08}; \citealt{jacob17}; \citealt{ruszkowski17}; \citealt{ehlert18}). 
 
The study of outbursts in a background medium has a long history in fluid dynamics (\citealt{landaufluid}, \citealt{thorneblandford}) and astrophysics, especially in supernova studies (\citealt{kingbook07}, \citealt{tang05}). As a strong outburst happens, the ambient medium is shocked and swept up while a reverse shock forms within the contact discontinuity between the ejecta and the shocked ambient medium. During the early stage when the mass swept up by the shock is much less than the ejecta mass, the outburst behaves as free expansion. When the swept-up mass becomes comparable to or larger than the ejecta mass, the outburst goes into the classic Sedov-Taylor phase. During this phase, the system is mainly controlled by the outburst energy and the background density, while the initial energy in the ambient medium within the shock front is negligible compared to the outburst energy. As the shock front detaches away from the contact discontinuity, the initial energy of the ambient medium within the shock front becomes dynamically important, and the system goes into a wave-like phase as described in \citet[also see \citealt{clarkebook07}; \citealt{tang05}]{tang17}. In observations of galaxy clusters (\citealt{mcnamara07}; \citealt{mcnamara12}), AGN jet ejecta contain magnetic fields, cosmic rays, and potentially very high-temperature low-density gas, and are usually identified as bright radio lobes or X-ray cavities enclosed by the shocked ICM. 

An important topic related to the complex energy transport and dissipation processes in AGN feedback is the energy partition between AGN ejecta and the ICM, i.e., the fraction of AGN jet energy transferred to the ICM, which is often denoted as the energy coupling efficiency $\eta_{\rm cp}$. This topic has been previously studied in some hydrodynamic simulations (\citealt{zanni05}; \citealt{binney07}; \citealt{weinberger17}; \citealt{english19}), which however usually focus on the hydrodynamic processes during the jet evolution, the heating mechanisms or observational signatures of AGN feedback in the ICM. Here in this paper we present a systematic study on the energy coupling efficiency of AGN outbursts in the ICM. In section \ref{section2}, we first estimate the values of $\eta_{\rm cp}$ in two extreme situations: quasi-static isobaric outbursts (the ``slow piston" limit for very mild outbursts) and very powerful outbursts going through the classic Sedov-Taylor phase. To get further physical insights, in section \ref{section3} we investigate the energy coupling efficiency of spherical outbursts in a uniform medium with a series of hydrodynamic simulations and perform a large parameter study over the total energy, outburst duration, thermal energy fraction, Mach number of the outbursts. In section \ref{section4}, we investigate the energy coupling efficiency of AGN jet outbursts in a realistic ICM with a series of hydrodynamic simulations and perform a parameter study over the same large parameter space as for spherical outbursts. Finally, in section \ref{section5} we present a summary of our results with some brief discussions.


\section{Analytical Estimations}
\label{section2}

In this section, we consider two extreme cases of outbursts in a uniform medium: quasi-static isobaric outbursts and instantaneous point outbursts, where the energy coupling efficiency can be estimated analytically. Roughly speaking, for a given outburst energy in a uniform medium, these two cases refer to very weak and very powerful outbursts, respectively. In section \ref{section2.3}, we describe a characteristic outburst power for a given outburst energy in a uniform medium that roughly separates these two extreme cases.

\subsection{Quasi-static Isobaric Outbursts}
\label{section2.1}

The first case refers to a very mild outburst which injects thermal gas into a uniform medium very slowly. The ejecta's expansion is approximately quasi-static, and isobaric with respect to the ambient medium. In this case, the energy coupling efficiency is the ratio of the pdV work done by the expanding ejecta to the enthalpy of the ejecta:
\begin{eqnarray} 
\eta_{\rm cp}=\frac{pV}{H}
\end{eqnarray}
where p, V and $H=\frac{\gamma}{\gamma-1}pV$ are the pressure, volume and enthalpy of the expanding ejecta bubble respectively. Thus $\eta_{\rm cp}=(\gamma-1)/\gamma$, and $\eta_{\rm cp}=0.4$ for $\gamma=5/3$. Note that the quasi-static expansion is a reversible process and does not replenish the entropy of the ambient gas lost through radiative cooling. As a comparison, shocks induced by powerful outbursts increase the entropy of the ambient gas. The isobaric outburst is ideal, and as a result of the second law of thermodynamics, the entropy of a real physical system would always increase due to unavoidable dissipation processes even for very mild outbursts. 

\subsection{Instantaneous Point Outbursts}
\label{section2.2}

For a strong outburst in the Sedov-Taylor phase, the surrounding medium with mass comparable to or larger than the injected gas mass is being swept up by a strong forward shock. In the shock frame, the pressure, density and velocity jump conditions of a strong shock with a very high Mach number can be written as follows, 
\begin{eqnarray}
P_{2}=  \frac{2}{ \gamma +1}  \rho_{0}v_{\rm s}^{2} \text{,}      \label{eqP2} \\
\rho_{2}=  \frac{\gamma+1}{ \gamma -1}  \rho_{0} \text{,}\\
v_{2}=  \frac{\gamma-1}{ \gamma +1}  v_{\rm s} ,
\end{eqnarray}
where the subscripts `0' refer to the variables in the shock upstream and '2' for the downstream. Here $v_{\rm s} = \frac{dR}{dt}$ is the the propagation velocity of the shock front in the upstream frame, and $R$ is the distance of the shock front from the origin where a spherical outburst is injected. We use $v_{\rm s}$ to replace $v_{0}$ since they have the same magnitudes. Then the velocity of the downstream postshock gas in the upstream frame is  
\begin{eqnarray}
v_{\rm ps} =v_{\rm s}-v_{2}= \frac{2}{\gamma+1} v_{\rm s} .
\end{eqnarray}

In the Sedov-Taylor phase, the shocked ambient gas is concentrated in a thin shell right behind the shock front. Thus one may adopt the thin shell approximation, and the mass and kinetic energy of the shocked ambient gas are dominated by the thin shell swept up by the shock front. The mass of the thin shell can be written as $M_{\rm sw}(t)  \approx  \frac{4 \pi }{3} \rho_{0}R^{3}$ and therefore its kinetic energy is $E_{\rm k} \approx \frac{2\pi}{3}\rho_{0}R^{3}v_{\rm ps}^{2}\propto \rho_{0}R^{3}v_{\rm s}^{2}$. For an outburst in the Sedov-Taylor phase, it has been shown that a simple linear relation exists between the kinetic energy of the swept-up gas and the outburst energy, which can be written as (\citealt{clarkebook07}; \citealt{achterbergbook})
\begin{eqnarray}  
E=\alpha_{\gamma} \rho_{0} R^3 v_{\rm s}^2 ,
\label{eqE}
\end{eqnarray}
where $\alpha_{\gamma}$ is a function of the adiabatic index $\gamma$ and also depends on the outburst history, e.g., an instantaneous point outburst or a strong but continuous outburst. For self-similar Sedov-Taylor solutions (e.g., \citealt{ostriker88}), one has $R\propto t^{\beta}$, and thus Equation (\ref{eqE}) can be rewritten as  
\begin{eqnarray}  
E=\kappa_{\gamma} \rho_{0}\frac{R^5}{t^2}.
\end{eqnarray}
This is the Sedov-Taylor relation originally derived via dimensional analysis (\citealt{choudhuribook}; \citealt{thorneblandford}). For an instantaneous point outburst, one has $\beta=2/5$ and thus $\kappa_{\gamma}=\frac{4}{25} \alpha_{\gamma}$. For a strong continuous outburst with a constant power, one has $\beta=3/5$ and thus $\kappa_{\gamma}=\frac{9}{25} \alpha_{\gamma}$. 

To derive the energy coupling efficiency, we should compare the total injected outburst energy with the outburst energy that has been transferred to the ambient medium. Since in the Sedov-Taylor phase the original energy of the shocked ambient gas is negligible compared to the outburst energy, the total energy in the shocked shell roughly equals the outburst energy transferred to it. The kinetic and thermal energies in the thin shell can be estimated as follows,
\begin{eqnarray}  
E_{\rm k,\rm sh}  \approx \frac{1}{2} \rho_{2} v_{\rm ps}^{2}  \cdot 4 \pi R^{2} \Delta R \text{,}\\
E_{\rm th,\rm sh}  \approx \frac{P_{2}}{\gamma -1} \cdot 4 \pi R^{2} \Delta R \text{,}
\end{eqnarray}
where $\Delta R$ is the thickness of the shell. Substituting $\rho_{2}$ and $P_{2}$ from the shock jump conditions, the kinetic and thermal energies can be written in the same form
\begin{eqnarray} 
E_{\rm k,\rm sh}\text{,~} E_{\rm th,\rm sh} \approx \frac{2\rho_{0}v_{\rm s}^2}{\gamma^2 -1} \cdot 4 \pi R^{2} \Delta R , 
\end{eqnarray}
which implies that in the thin shell approximation there is an equipartition between the kinetic and thermal energies of the shocked gas shell. Then by equaling two forms of the shell mass $\frac{4 \pi }{3} \rho_{0}R^{3}=4 \pi \rho_{2} R^{2} \Delta R $, one can derive the thickness of the shell 
\begin{eqnarray}  
\frac{\Delta R}{R}=\frac{\gamma-1}{3(\gamma+1)} ,
\end{eqnarray}
which is equal to $0.08$ and indeed very small for $\gamma = 5/3$. Then recalling the form of the outburst energy $E$ in Equation (\ref{eqE}), the total energy of the shocked shell can be written as 
\begin{eqnarray}  
E_{\rm shell} =E_{\rm k,\rm sh}+E_{\rm th,\rm sh} \approx \frac{16 \pi}{3 \alpha_{\gamma} (\gamma +1)^2}  E .
\end{eqnarray}

We define the energy coupling efficiency $\eta_{\rm cp}$ as the fraction of the outburst energy transferred to the ambient medium, which can be estimated as $E_{\rm shell}/E$. For an instantaneous point outburst, $\kappa_{\gamma}=\frac{4}{25} \alpha_{\gamma}$ as we derive above and then     
\begin{eqnarray} 
 \eta_{\rm cp} \approx  \frac{64 \pi}{75 \kappa_{\gamma} (\gamma +1)^2} \approx 0.77 ,
\end{eqnarray}
where we take $\kappa_{\gamma} \approx 0.49$ for $\gamma=\frac{5}{3}$ from previous studies of the self-similar Sedov-Taylor solution of instantaneous outbursts (\citealt{taylor50}; \citealt{petruk00}; \citealt{thorneblandford}). After the Sedov-Taylor phase, the shock front detaches away from the contact discontinuity, and the outburst ejecta continues to expand before halting, leading to more energies transferred to the ambient medium. Thus the energy coupling efficiency for a real strong outburst is expected to be larger than the value estimated here $\eta_{\rm cp}\gtrsim 0.77$, as further confirmed by hydrodynamic simulations in Section 3.2.    

\begin{deluxetable*}{lccccccccccc}
\label{tab1}
\tablenum{1}
\tablecaption{List of Our Spherical Outburst Simulations}
\tablewidth{0pt}
\tablehead{
\colhead{Run} & \colhead{$E_{\rm inj}$} &  \colhead{$t_{\rm inj}$} & \colhead{$f_{\rm th} $}  & \colhead{ $M_{\rm inj}$}  &\colhead {$p_{\rm inj} /p_{0}$}  & \colhead{$\rho_{\rm inj}/\rho_{0}$} & \colhead{$p_{\rm tot,\rm inj}/p_{0}$}  & \colhead{$\eta_{\rm cp}$ } & \colhead{$r_{\rm stop}$} & \colhead{$r_{\rm sonic}$ } & \colhead{$t_{\rm sonic}$}
}
\startdata
W       &     2.5        &   0.1       &   0.1	&     5          &  $4.42\times10^{2}$ & $2.86\times10^{2}$  & $7.59\times10^{3}$ & 0.85           & 0.46         & 0.81           &0.24    \\
Wt0    &      ---        &   0.05     &    ---         &    ---          &  $8.84\times10^{2}$ & $5.73\times10^{2}$  & $2.47\times10^{4}$ & 0.92           & 0.40         & 0.86           &0.29    \\
Wt1    &      ---        &   0.2       &    ---         &    ---          &  $2.21\times10^{2}$ & $1.43\times10^{2}$  & $3.80\times10^{3}$ & 0.78           & 0.57         & 0.60           &0.22    \\
Wt2    &      ---        &   0.5       &    ---         &    ---          &  $8.84\times10^{1}$ & $5.73\times10^{1}$  & $2.47\times10^{3}$ & 0.63           & 0.65         & 0.37           &0.14    \\
Wt3      &     ---        &   10        &   ---	      &     ---          &  $4.42$ & $2.86$  & $7.59\times10^{1}$ & 0.42           & 0.77         & 0.08           &0.03    \\
Wf1    &      ---        &    ---       &   0.9         &    ---          &  $3.89\times10^{3}$ & $3.18\times10^{1}$  & $5.30\times10^{3}$ & 0.82           & 0.57         & 0.85           &0.28    \\
WM1  &      ---        &    ---       &    ---         &    20          &  $1.11\times10^{2}$ & $4.48\times10^{0}$  & $3.10\times10^{3}$ & 0.82           & 0.55         & 0.79           &0.26    \\
WE1   &     0.25     &    ---       &    ---         &    ---          &  $4.42\times10^{1}$ & $2.86\times10^{1}$  & $1.23\times10^{3}$ & 0.74           & 0.25         & 0.27           &0.1    \\
\enddata
\tablecomments{The parameters in our spherical outburst simulations include the outburst energy $E_{\rm inj}$, duration $t_{\rm inj}$, thermal fraction $f_{\rm th}$ and Mach number $M_{\rm inj}$. These parameters determine the ejecta properties at the inner boundary (base): pressure contrast $p_{\rm inj} /p_{0}$ with respect to the ambient gas pressure, density contrast $\rho_{\rm inj}/\rho_{0}$ with respect to the ambient gas density, and the total pressure ratio $p_{\rm tot,\rm inj}/p_{0}$, where $p_{\rm tot,\rm inj}= p_{\rm inj} +\rho_{\rm inj}M_{\rm inj}^2c_{\rm s0}^2$ includes both thermal and ram pressures. The mark '---' means that the corresponding parameter has the same value as in the fiducial run (run W).  $\eta_{\rm cp}$ is the energy coupling efficiency of the outburst in the ambient medium, and $r_{\rm stop}$ is the final radius of the ejecta bubble, both measured at $t=t_{\rm inj}+t_{\rm s}$, where $t_{\rm s}$ is a characteristic outburst timescale defined in Section 2.3. The outburst drives a forward shock into the ambient medium. The sonic radius $r_{\rm sonic}$ is the location of the forward shock at $t=t_{\rm sonic}$, which is defined as the time when the postshock gas velocity becomes equal to the ambient sound speed.
}   
\end{deluxetable*}
 
\subsection{A Characteristic Outburst Power}
\label{section2.3}

For a given outburst energy $E_{\rm inj}$ injected in a uniform background with uniform density $\rho_{0}$, and pressure $p_{0}$, there exists a characteristic outburst power that roughly separates the two extreme cases of weak and powerful outbursts described above. In the Sedov-Taylor approximation, the original thermal energy of the swept-up ambient medium is much less than the injected outburst energy. In other words, the outburst energy $E_{\rm inj}$ defines a characteristic feedback radius $R_{\rm fb}$ within which the initial thermal energy of the ambient gas equals $E_{\rm inj}$:
 \begin{eqnarray} 
 E_{\rm inj}=\frac{4\pi }{3}R_{\rm fb}^{3}p_{0}.
 \label{Efb}
\end{eqnarray}
The above equation may also be interpreted as the outburst induces pressure perturbations $\delta p$ in the ambient medium, and within $R_{\rm fb}$, $\delta p$ is large enough to be comparable to $p_{0}$ and the total energy stored in the perturbations is comparable to $E_{\rm inj}$, i.e., $E_{\rm inj}=  \int _ {0} ^ { R_{\rm fb} } 4 \pi r^2 \delta p dr$.

A characteristic timescale of the outburst may be defined as the sound crossing time across the feedback radius $R_{\rm fb}$ 
 \begin{eqnarray} 
 t_{\rm s}=\frac{R_{\rm fb}}{c_{\rm s0}} , 
\end{eqnarray}
where $c_{\rm s0}=\sqrt{\gamma p_{0}/\rho_{0}}$. Note that the same definitions of $R_{\rm fb}$ and $t_{\rm s}$ have been previously proposed in \citet{tang17}. With $R_{\rm fb}$ and $t_{\rm s}$, one can define a characteristic outburst power
\begin{eqnarray} 
P_{\rm fb}=E_{\rm inj}/t_{\rm s}. 
\end{eqnarray}
For a given outburst energy $E_{\rm inj}$, an outburst with power $P\gg P_{\rm fb}$ can be considered as an instantaneous powerful outburst with the energy coupling efficiency $\eta_{\rm cp}\gtrsim 0.77$, while an outburst with $P\ll P_{\rm fb}$ may be approximated as a slow isobaric outburst with $\eta_{\rm cp}\sim0.4$. We will investigate and confirm these results with hydrodynamic simulations in Section \ref{section3}. The above argument also suggests that in a real system, an outburst with energy $E_{\rm inj}$ is potentially important in heating the ambient medium within a radius of $R_{\rm fb}$ during a timescale of $t_{\rm s}$, and to effectively offset radiative cooling, $R_{\rm fb}$ should be comparable to or larger than the system's cooling radius, within which the gas cooling time is shorter than the system's age. 
 

\begin{figure*}
\centering
\includegraphics[height=0.27\textheight]{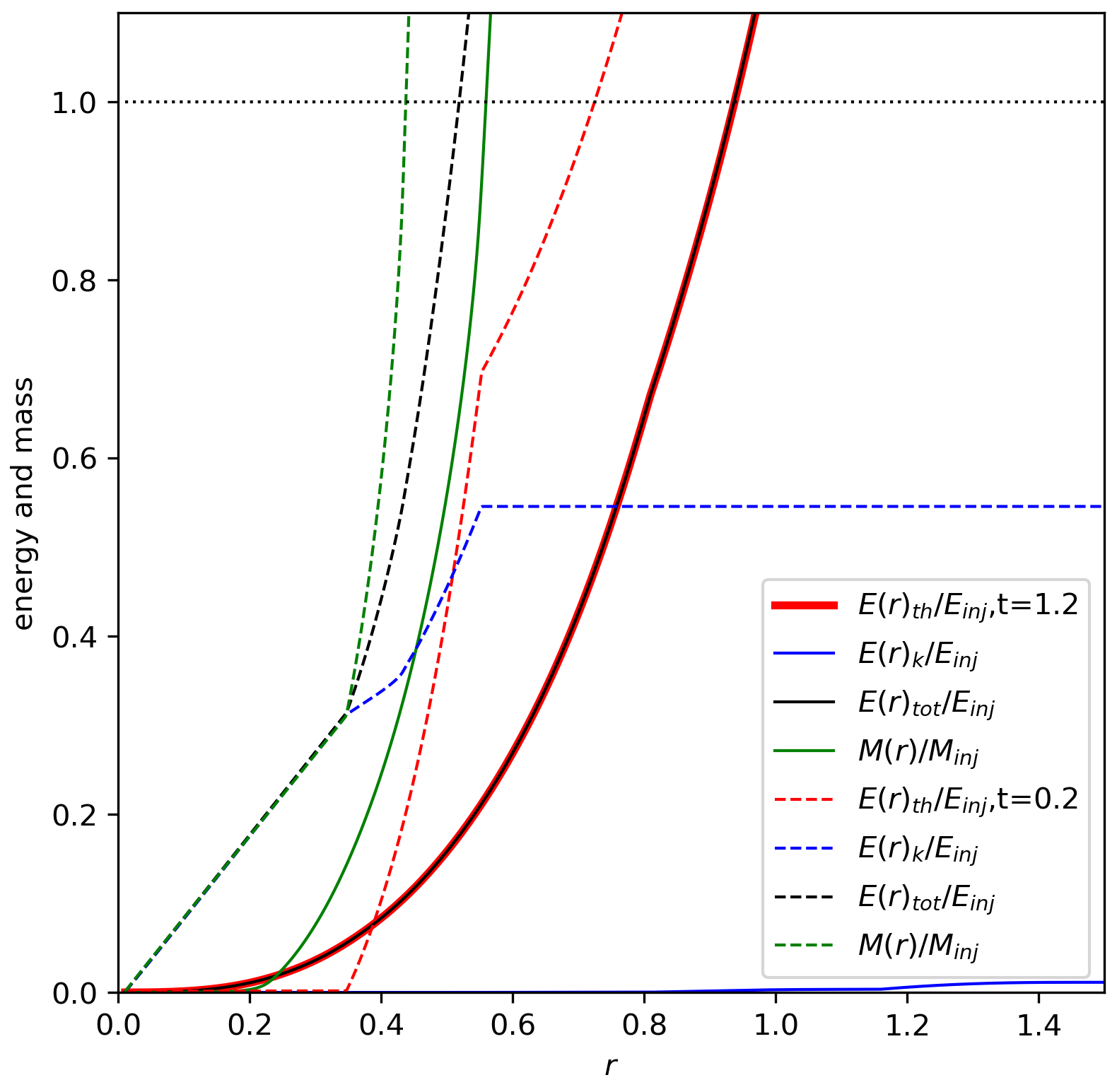}~~~~~~~~~~
\includegraphics[height=0.27\textheight]{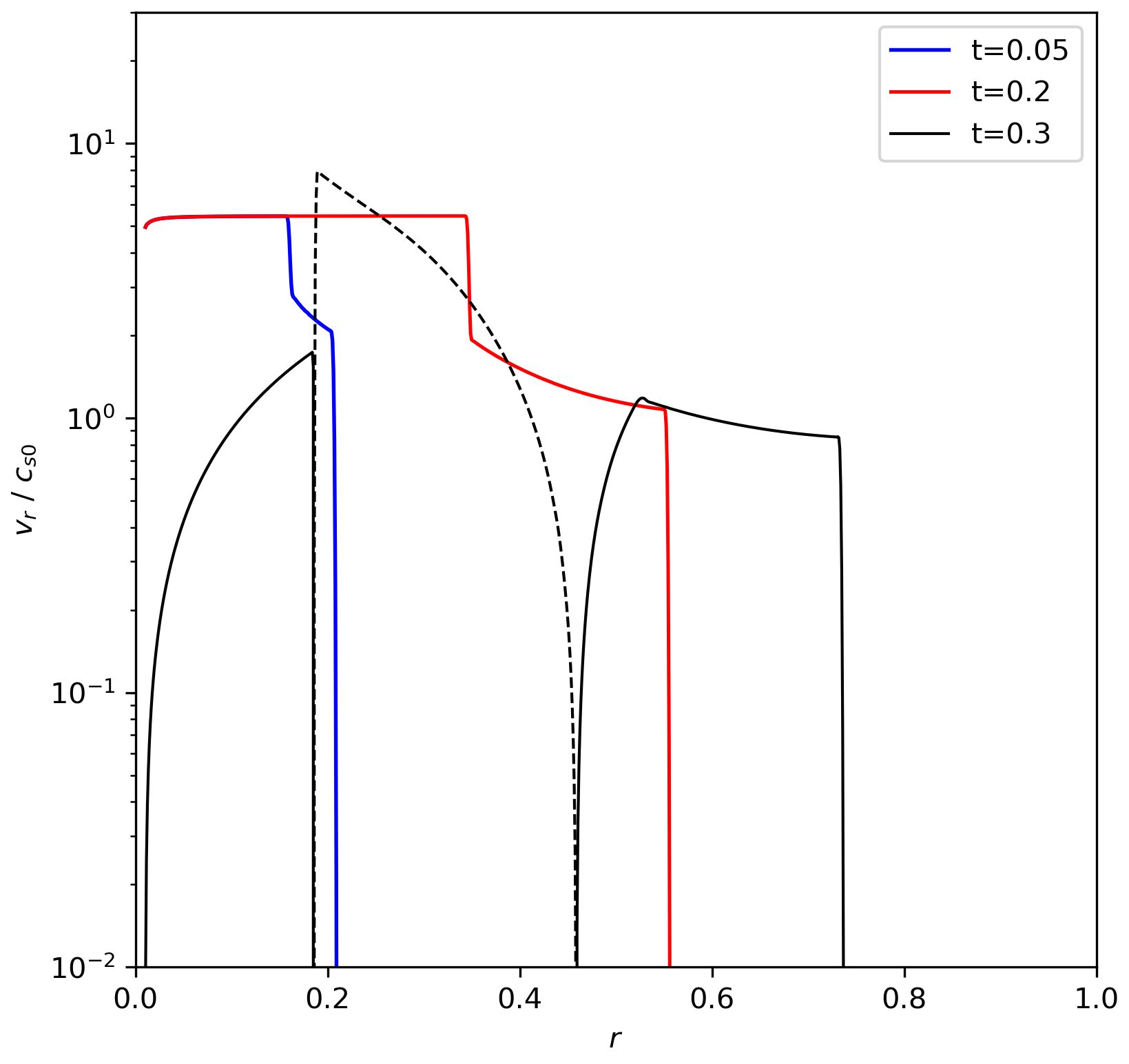}
\caption{A representative spherical outburst in run Wt1 with $P_{\rm inj}=5P_{\rm fb}$. {\it Left}: the radial profiles of integrated gas mass and energies within radius $r$, normalized by the total injected mass $M_{\rm inj}$ and energy $E_{\rm inj}$, respectively. The subscripts ``k", ``th", and ``tot" refer to the kinetic, thermal, and total energies, respectively. The dashed lines refer to the profiles at $t=0.2$ right when the outburst ends, while the solid lines are for $t=1.2$, a sound crossing time across $R_{\rm fb}$ after the outburst ends, when the size of the ejecta bubble is stable. Note that the black solid line overlays on the red solid one. The crossing of the profile of $M(r)/M_{\rm inj}$ with the horizontal dotted line marks the size of the ejecta bubble. {\it Right}: the radial velocity profiles at $t=0.05, 0.2, 0.3$, normalized by the sound speed of the ambient medium $c_{\rm s0}$. The dashed line stands for negative values of radial velocity.}
\label{plot1}
\end{figure*}

\section{Spherical Outbursts in a Uniform Medium}
\label{section3}

In this section, to gain further physical insights, we investigate spherical outbursts in a uniform medium with a series of one-dimensional (1D) hydrodynamic simulations under the assumption of spherical symmetry.

\subsection{Numerical Setup}
\label{section3.1}

For simplicity, we solve the non-dimensionalized hydrodynamic equations and neglect gravity, radiative cooling and viscosity. The simulations were performed in spherical coordinates using the 1D mode of the code ZEUS-MP (\citealt{zeusmp}). The radial computational domain extends from an inner boundary of $r_{\rm in}=0.01$ to an outer boundary of $2.0$ with 2000 uniform grids. At the outer boundary, we use the outflow boundary conditions, and at the inner boundary, we use the reflecting boundary conditions except during the outburst $0 \leq t \leq t_{\rm inj}$, when thermal gas is injected across $r_{\rm in}$ using the inflow boundary conditions with a constant power $P_{\rm inj}$. 

The outbursts in our simulations carry both thermal and kinetic energies. At the inner boundary, we set up an outburst with four parameters: the total outburst energy $E_{\rm inj}$, duration $t_{\rm inj}$, thermal fraction $f_{\rm th}$ as the ratio of the injected thermal energy to $E_{\rm inj}$, and the Mach number $M_{\rm inj}$ as the ratio of the ejecta speed to the sound speed of the ambient medium. The values of $E_{\rm inj}$, $f_{\rm th}$, and $M_{\rm inj}$ determine the gas density $\rho_{\rm inj}$, pressure $p_{\rm inj}$, and velocity $v_{\rm inj}$ of the outburst ejecta at the inner boundary. We performed a large suite of simulations of spherical outbursts over the parameter space ($E_{\rm inj}, t_{\rm inj}, f_{\rm th}, M_{\rm inj}$). The input parameters, ejecta properties and main results of some representative simulations are listed in Table 1. 

In our simulations, we adopt a uniform background medium with density $\rho_{0}=1$ and pressure $p_{0}=0.6$. The adiabatic sound speed in the ambient medium is thus $c_{\rm s0}=(\gamma p_{0}/\rho_{0})^{1/2}=1$ for $\gamma=5/3$. In most of our simulations (except run WE1), the feedback radius is chosen to be $R_{\rm tb}=1$, corresponding to an outburst energy of $E_{\rm inj}\approx 2.5$ according to Equation (\ref{Efb}). The characteristic timescale is thus $t_{\rm s}\equiv R_{\rm fb}/c_{\rm s0}=1$. The value of $t_{\rm s}$ sets the baseline values for the outburst duration $t_{\rm inj}$ and the outburst power $P_{\rm inj}$. In our fiducial study (run W listed in Table 1), the outburst duration is chosen to be $t_{\rm inj}=0.1$, corresponding to a very powerful outburst with $P_{\rm inj}=10P_{\rm fb}$. The Mach number at the base is chosen to be $M_{\rm inj}=5$.

\subsection{Results}
\label{section3.2}

\begin{figure}
\includegraphics[height=0.27\textheight]{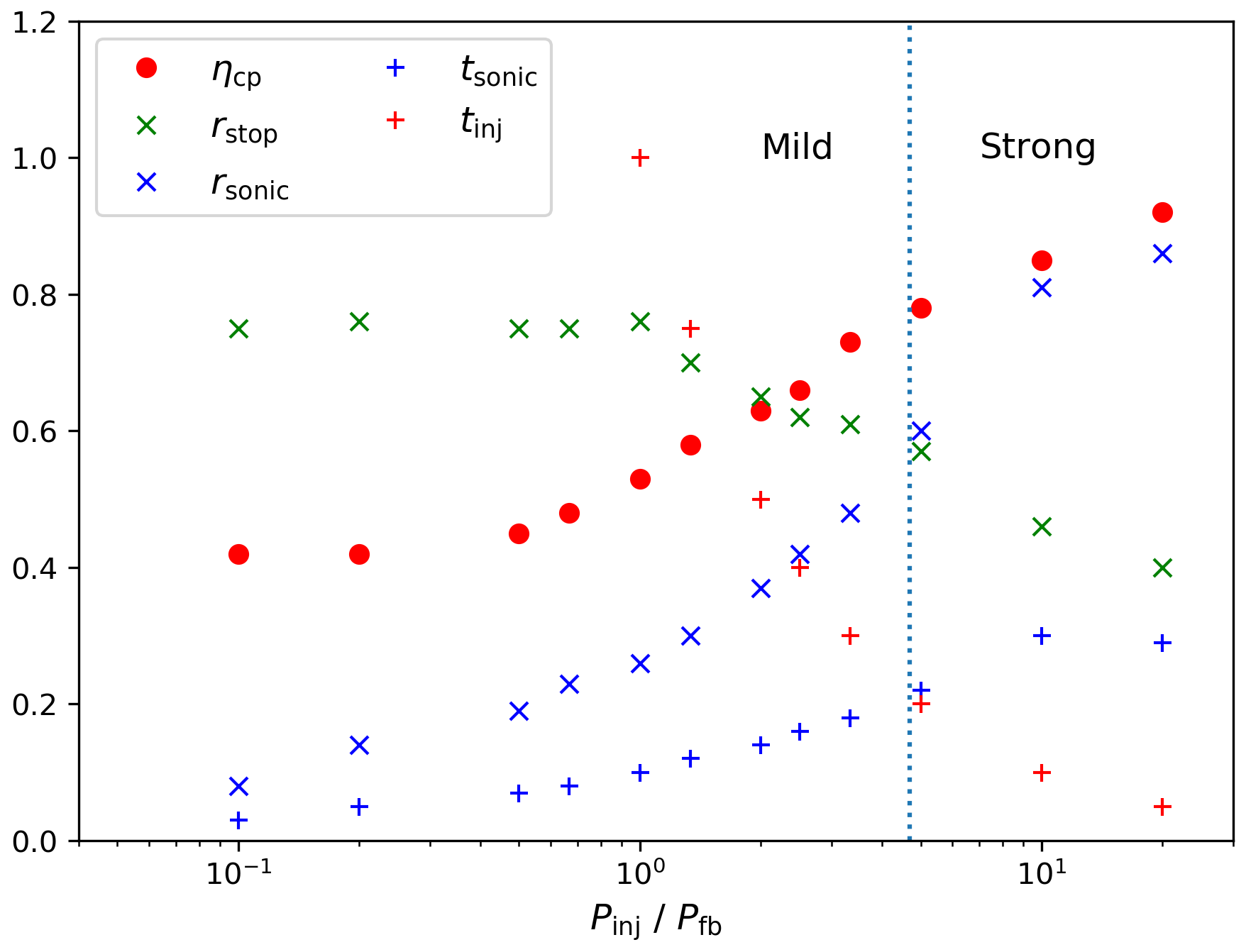}
\caption{Some key results of our spherical outburst simulations, where all the model parameters are the same as in run W except the outburst duration $t_{\rm inj}$. The outburst power is related with $t_{\rm inj}$ through $P_{\rm inj}/P_{\rm fb}=t_{\rm s}/t_{\rm inj}$, where the characteristic outburst power $P_{\rm fb}$ and timescale $t_{\rm s}$ for a given outburst energy are defined in Section \ref{section2.3}. The vertical dotted line denotes the value of $P_{\rm inj}/P_{\rm fb}$ when $t_{\rm sonic}=t_{\rm inj}$. }
\label{plot2}
\end{figure}

In each of our simulations, we identify the boundary of the ejecta bubble at every time step as the radius $r_{\rm ejecta}$ within which the total gas mass is equal to the injected mass. We then evaluate the total energy in the ejecta bubble $E(r_{\rm ejecta})$, and thus determine the energy coupling efficiency of the outburst with the ambient medium as $\eta_{\rm cp}=1- E(r_{\rm ejecta})/E_{\rm inj}$. In our simulations, the ejecta bubble expands during the outburst, and gradually stops expansion after the outburst ends with $r_{\rm ejecta}=r_{\rm stop}$. The values of $\eta_{\rm cp}$ and the final radius of the ejecta bubble $r_{\rm stop}$ listed in Table 1 are evaluated at $t=t_{\rm inj}+t_{\rm s}$, when the size of the ejecta bubble has already been stable. The left panel of Figure 1 shows the radial profiles of the integrated mass and energies in a representative run (run Wt1) at $t=t_{\rm inj}$ and $t=t_{\rm inj}+t_{\rm s}$, while the right panel shows the radial velocity profiles at three times $t=0.05$, $0.2$, and $0.3$. 

Comparing other runs in Table 1 with run W, one can see that for a given outburst energy $E_{\rm inj}$, the energy coupling efficiency $\eta_{\rm cp}$ mainly depends on the outburst duration $t_{\rm inj}$, i.e., the outburst power $P_{\rm inj}=E_{\rm inj}/t_{\rm inj}$, and is rather insensitive to other outburst parameters such as the Mach number $M_{\rm inj}$ and the thermal fraction $f_{\rm th}$. Figure 2 shows some key results, including $\eta_{\rm cp}$, of our spherical outburst simulations, where all the model parameters are the same as in run W except the outburst duration $t_{\rm inj}$. The outburst power in each run is related with $t_{\rm inj}$ through $P_{\rm inj}/P_{\rm fb}=t_{\rm s}/t_{\rm inj}$. It is clear that $\eta_{\rm cp} \sim 0.4$ when $P_{\rm inj}/P_{\rm fb} \ll 1$ and $\eta_{\rm cp} \gtrsim 0.77$ when $P_{\rm inj}/P_{\rm fb} \gg 1$, confirming our analytical estimates of $\eta_{\rm cp}$ in quasi-static isobaric outbursts and instantaneous point outbursts in Section 2, respectively. 

Another transitioning power $P_{\rm tr}\sim 5P_{\rm fb}$ for spherical outbursts can also be seen in Figure 2, which shows the dependence of $\eta_{\rm cp}$, $t_{\rm inj}$, $r_{\rm stop}$, $t_{\rm sonic}$, and $r_{\rm sonic}$ on the outburst power. The outburst drives a forward shock propagating into the ambient medium. The sonic time $t_{\rm sonic}$ here is defined as the time when the velocity of the postshock gas at the shock front becomes equal to the sound speed in the ambient gas. The sonic radius $r_{\rm sonic}$ is the distance of the forward shock to the origin at $t=t_{\rm sonic}$. As illustrated in Figure 2, $t_{\rm sonic}=t_{\rm inj}$ when $P_{\rm tr}\sim 5P_{\rm fb}$. In other words, the postshock gas becomes subsonic with respect to the ambient gas before the outburst ends if $P_{\rm inj}<5P_{\rm fb}$, or equivalently $t_{\rm inj} > 0.2t_{\rm s}$. The gas velocity profiles in run Wt1 where $P_{\rm inj}=P_{\rm tr}$ before and after $t_{\rm inj}$ are shown in the right panel of Figure 1, confirming that the postshock gas velocity in this run is indeed roughly equal to the ambient sound speed at $t\sim t_{\rm inj}=0.2$. Figure 2 also shows that interestingly, the final radius of the ejecta bubble $r_{\rm stop}$ roughly equals $r_{\rm sonic}$ ($r_{\rm stop}\approx r_{\rm sonic}\approx 0.6R_{\rm fb}$) when $P_{\rm inj}=P_{\rm tr}$. More importantly, for outbursts with $P_{\rm inj}>P_{\rm tr}$, $\eta_{\rm cp} \gtrsim 0.8$ which is about the value estimated analytically in Section 2.2 for instantaneous point outbursts in the thin shell approximation. 

The transitioning power $P_{\rm tr} \approx 5P_{\rm fb}$ (or equivalently the transitioning outburst duration $t_{\rm tr} \approx 0.2t_{\rm s}$) is general for any given outburst energy in a uniform background. In run WE1 we simulate the evolution of an outburst with $E_{\rm inj}=0.25$, which is ten times smaller than in run W. As $t_{\rm s}\varpropto E_{\rm inj}^{1/3}$, we have $t_{\rm s}\approx 0.46$ and $t_{\rm tr} \approx 0.09$. In run WE1, we choose $t_{\rm inj}=0.1 \approx t_{\rm tr}$, resulting in $t_{\rm sonic} = t_{\rm inj}\approx0.2t_{\rm s}$ and $r_{\rm sonic} \approx r_{\rm stop} \approx 0.6 R_{\rm fb}$, as listed in Table 1. Here $R_{\rm fb}=0.46$ as $R_{\rm fb}\varpropto E_{\rm inj}^{1/3}$. A similar transitioning duration $t_{\rm tr} \approx 0.15t_{\rm s}$ has also been suggested in \citet{tang17} and \citet{tang18} in view of energy partition which show that the shock-heated shell captures the majority of the outburst energy when $t_{\rm inj} < 0.15 t_{\rm s}$, consistent with our results here.       

\begin{deluxetable*}{lcccccccclcc}
\label{tab2}
\tablenum{2}
\tablecaption{List of Our AGN Jet Simulations}
\tablewidth{0pt}
\tablehead{
\colhead{} & \colhead{$E_{\rm inj}$}&  \colhead{$t_{\rm inj}$} & \colhead{$f_{\rm th} $}  & \colhead{ $M_{\rm inj}$}& \colhead{ $P_{\rm inj}$}  &\colhead {$p_{\rm inj} /p_{0}$}  & \colhead{$\rho_{\rm inj}/\rho_{0}$} & \colhead{$p_{\rm tot,\rm inj}/p_{0}$}  & \colhead{$\eta_{\rm cp}$ } &  \colhead{$t_{\rm cc}$} \\
\colhead{Run} & \colhead{$(10^{60}\rm erg)$}&  \colhead{$(\rm Myr)$} & \colhead{}  & \colhead{}& \colhead{$(10^{45}\rm erg\,\rm s^{-1})$}  &\colhead {}  & \colhead{} & \colhead{}  & \colhead{} &  \colhead{$(\rm Myr)$} 
}
\startdata
J       &     2.3        &   5          &   0.1         &     35         &     14.6         &  8.91        & 0.12                             & 169.22        & 0.88          & 477   \\
Jt1    &      ---        &   50        &    ---          &    ---         &    1.46          &  0.89        & 0.01                             &  16.92         & 0.77          & 390    \\
Jt2    &      ---        &   100      &    ---          &    ---         &    0.73          &  0.45        & $5.9 \times10^{-3}$     &  8.47           & 0.83          & 384   \\
Jt3    &      ---        &   300      &    ---          &    ---          &    0.24         &  0.15        & $2.0 \times10^{-3}$     &  2.82           & 0.86          & 383   \\
Jf0    &      ---        &    ---       &   0.0         &    ---           &    ---          &   0.00       &  0.13                            & 178.13        & 0.89          & 434    \\
Jf1    &      ---        &    ---       &   0.5         &    ---           &    ---          &  44.55       & 0.07                            & 133.61        & 0.85          &761    \\
Jf2    &      ---        &    ---       &   0.9         &    ---           &    ---          &  80.18       & 0.01                            &  98.00         & 0.83          & $>$800 \\
Jt1f0 &      ---        &    50        &  0.0         &   ---            &    1.46          &   0.00        &  0.01                           & 17.69          &0.81           & 391    \\
Jt1f1 &      ---        &    50        &  0.5         &   ---            &    1.46         &   4.45        & $6.5\times10^{-3}$     & 13.36           &0.73            & 437    \\
Jt1f2 &      ---        &    50        &  0.9         &   ---            &    1.46       &   8.02        & $1.0\times10^{-3}$     &  9.80            &0.68            & 450         \\
JE1   &    0.23       &    ---        &  ---          &   ---            &    1.46       &  0.89          &0.01                            & 16.92           &0.83            &391   \\
JE1t1&    0.23       &   50         &   ---         &   ---            &    0.15       &  0.09          &$1.2\times10^{-3}$     & 1.69             &0.77            &323       \\
JM1   &      ---         &   ---        &   ---          &    7             &    ---          & 44.55        & 14.73                         &  846.16        & 0.87           &179        \\
\enddata
\tablecomments{The parameters and physical quantities here are defined mostly the same as in Table 1, but for our jet simulations. $P_{\rm inj}$ and $E_{\rm inj}$ stand for the power and the total injected energy of one jet, respectively. The subscript `0' refers to the corresponding physical quantity in the initial ambient ICM at the jet base. The mark '---' means that the corresponding parameter has the same value as in the fiducial run (run J). In our default simulations, radiative cooling is neglected for simplicity. However, in Section 4.5, we rerun all the simulations with cooling included, and $t_{\rm cc}$ in the rightmost column refers to the start time of the central cooling catastrophe in these simulations.}
\end{deluxetable*}

\begin{figure*}
\centering
\includegraphics[height=0.25\textheight]{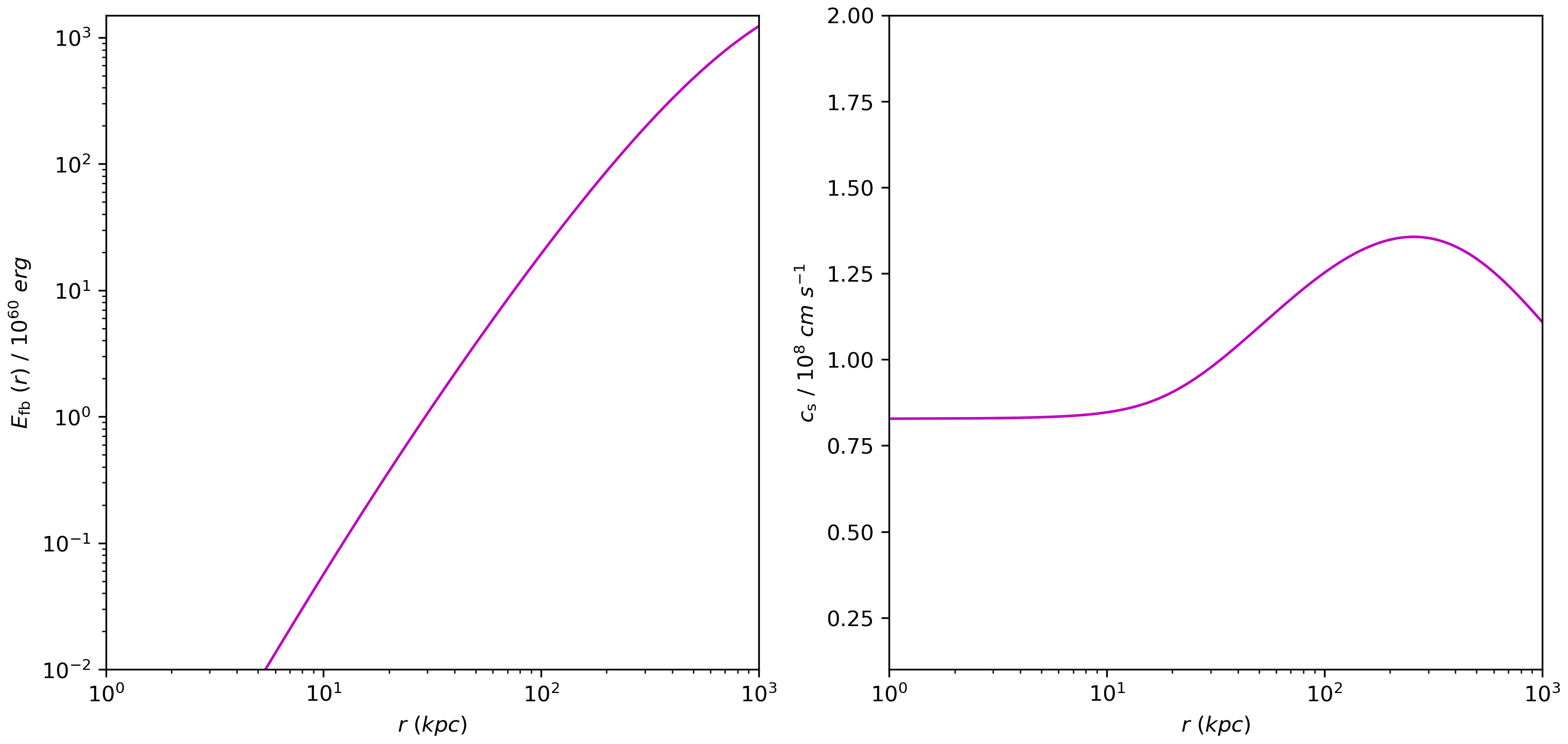}
\caption{Radial profiles of the characteristic AGN feedback energy calculated as $E_{\rm fb} (r)=\int_{0}^{r} 4 \pi r'^2 p(r') dr' $ and the sound speed in our simulated ICM environment. Here $p(r)$ is the initial pressure profile in the ICM.}
\label{plot3}
\end{figure*}

\section{AGN Jet Outbursts in a Realistic ICM}
\label{section4}

In this section, we present a large suite of hydrodynamic simulations to explore the energy coupling efficiency of AGN jet outbursts in a realistic ICM.

\subsection{Simulation Setup and the Characteristic Jet Powers}
\label{section4.1}

The setup of our jet simulations basically follows our previous work \citet{guo18} and \citet{duan18}, where we refer the readers for more details. Here we briefly describe the basic setup and some updates. We choose the well-observed galaxy cluster Abell 1795 as our model cluster. The ICM is initially in hydrostatic equilibrium in a static gravitational potential contributed by the dark matter halo, the central galaxy and supermassive black hole. The initial density and temperature distributions provide a very good fit to X-ray observations of this cluster. Assuming axisymmetry around the jet axis, we solve the basic hydrodynamic equations in $(R, z)$ cylindrical coordinates using a finite-difference Eulerian code. The computational grid along each axis consists of $800$ equally spaced zones with spatial resolution of $0.25$ kpc from the origin to $200$ kpc plus additional $400$ logarithmically-spaced zones out to $2$ Mpc. We use reflective boundary conditions at the inner boundaries and outflow boundary conditions at the outer boundaries. 

During the AGN outburst phase $0 \leq t \leq t_{\rm inj}$, a constant jet is injected along the $+z$ direction with a cross-section radius of $R_{\rm inj} =1.5$ kpc around the $z$ axis. We initialize the jet at $z=1$ kpc by adding gas fluxes of mass, momentum, and energy corresponding to a uniform jet with density $\rho_{\rm inj}$, energy density $e_{\rm inj}$ and velocity $v_{\rm inj}$. The total energy injected by one jet into our simulated domain is then $E_{\rm inj}=P_{\rm inj}t_{\rm inj}$. We performed a series of hydrodynamic jet simulations over the same outburst parameter space as in spherical outburst simulations, including $E_{\rm inj}$, $t_{\rm inj}$, thermal fraction $f_{\rm th}$ and the Mach number $M_{\rm inj}$ defined as the ratio of $v_{\rm inj}$ to the sound speed in the ambient ICM at the jet base. In our jet simulations, a relatively low level of shear viscosity with a dynamic viscosity coefficient $\eta_{\rm visc}=300$ g cm$^{-1}$ s$^{-1} $ is adopted to suppress the Kelvin-Helmholz instability (\citealt{reynolds05}; \citealt{guo12}; \citealt{guo15}; \citealt{duan18}). For simplicity, radiative cooling is neglected in all our jet simulations except that in Section 4.5, we rerun all the jet simulations with radiative cooling included to investigate how AGN outbursts with various jet parameters affect the development of the central cooling catastrophe.

As in Section \ref{section3.1}, a characteristic AGN feedback energy which is expected to significantly affect an ICM region with radius $R_{\rm fb}$ can be estimated as $E_{\rm fb}=\int _ {0} ^ { R_{\rm fb} } 4 \pi r^2 p(r) dr$, which is typically about several $10^{60}$ erg within several tens kpc. The radial profiles of $E_{\rm fb}$ and the sound speed $c_{\rm s0}$ in the initial ICM are shown in Figure 3. The sound speed in our ICM environment is typically about $10^8$ cm s$^{-1}$, and thus the sound crossing timescale within $50$ kpc is about 50 Myr:  
\begin{eqnarray} 
 t_{\rm s} \approx 50 \frac{R_{\rm fb}}{50\rm ~kpc} \left(\frac{c_{\rm s}}{10^{8}{\rm ~cm} {\rm ~s}^{-1}} \right)^{-1} {\rm ~Myr}.
\end{eqnarray} 
In most runs, we adopt the jet energy as $E_{\rm inj}=2.3\times 10^{60}$ erg, corresponding to a characteristic feedback radius of $R_{\rm fb}\approx 54$ kpc which is derived from $2E_{\rm inj}=\int _ {0} ^ { R_{\rm fb} } 4 \pi r^2 p(r) dr$ assuming two opposing jets for the outburst. The characteristic timescale within $R_{\rm fb}$ is roughly $t_{\rm s}=50$ Myr. We choose $t_{\rm inj}=0.1t_{\rm s}$ for typical strong jet outbursts, and $t_{\rm inj}=t_{\rm s}$ for typical mild outbursts. The jet power can be determined through $P_{\rm inj}=E_{\rm inj}/t_{\rm inj}$. We explore a large parameter space in our simulations, and the outburst parameters and main results are listed in Table 2.

As for spherical outbursts, one may expect that for a given outburst energy $2E_{\rm inj}$, there exist two characteristic jet powers $P_{\rm fb}=E_{\rm inj}/t_{\rm s}$ and $5P_{\rm fb}$, which roughly separate the two extreme cases of slow isobaric outbursts and instantaneous point outbursts. However, jet outbursts are not the same as spherical outbursts. The region affected by the former may be approximated as a spheroid instead of a sphere. Assuming that a jet outburst with energy $2E_{\rm inj}$ effectively affects a spheroid with a semi-major axis $z_{\rm fb}$ and two semi-minor axes $R_{\rm fb}$ in a uniform medium, one may have
 \begin{eqnarray} 
 2E_{\rm inj}=\frac{4\pi }{3}R_{\rm fb}^{2}z_{\rm fb}p_{0}=\frac{4\pi }{3}\alpha R_{\rm fb}^{3}p_{0},
\end{eqnarray}
where $\alpha=z_{\rm fb}/R_{\rm fb}$ may be considered as the aspect ratio of the induced forward shock. This would result in $t_{\rm s}\equiv R_{\rm fb}/c_{\rm s}\propto \alpha^{-1/3}$ and  $P_{\rm fb}\equiv E_{\rm inj}/t_{\rm s}\propto \alpha^{1/3}$. The value of $\alpha$ typically decreases with thermal fraction $f_{\rm th}$, and taking a typical value of a few for $\alpha$, the transitioning powers $P_{\rm fb}$ and $5P_{\rm fb}$ for a given outburst energy are expected to be slightly larger in jet outbursts than in spherical outbursts where $\alpha=1$.

\subsection{Ejecta and Shock Morphologies}
\label{section4.2}

\begin{figure*}
\centering
\gridline{
\includegraphics[height=0.55\textheight]{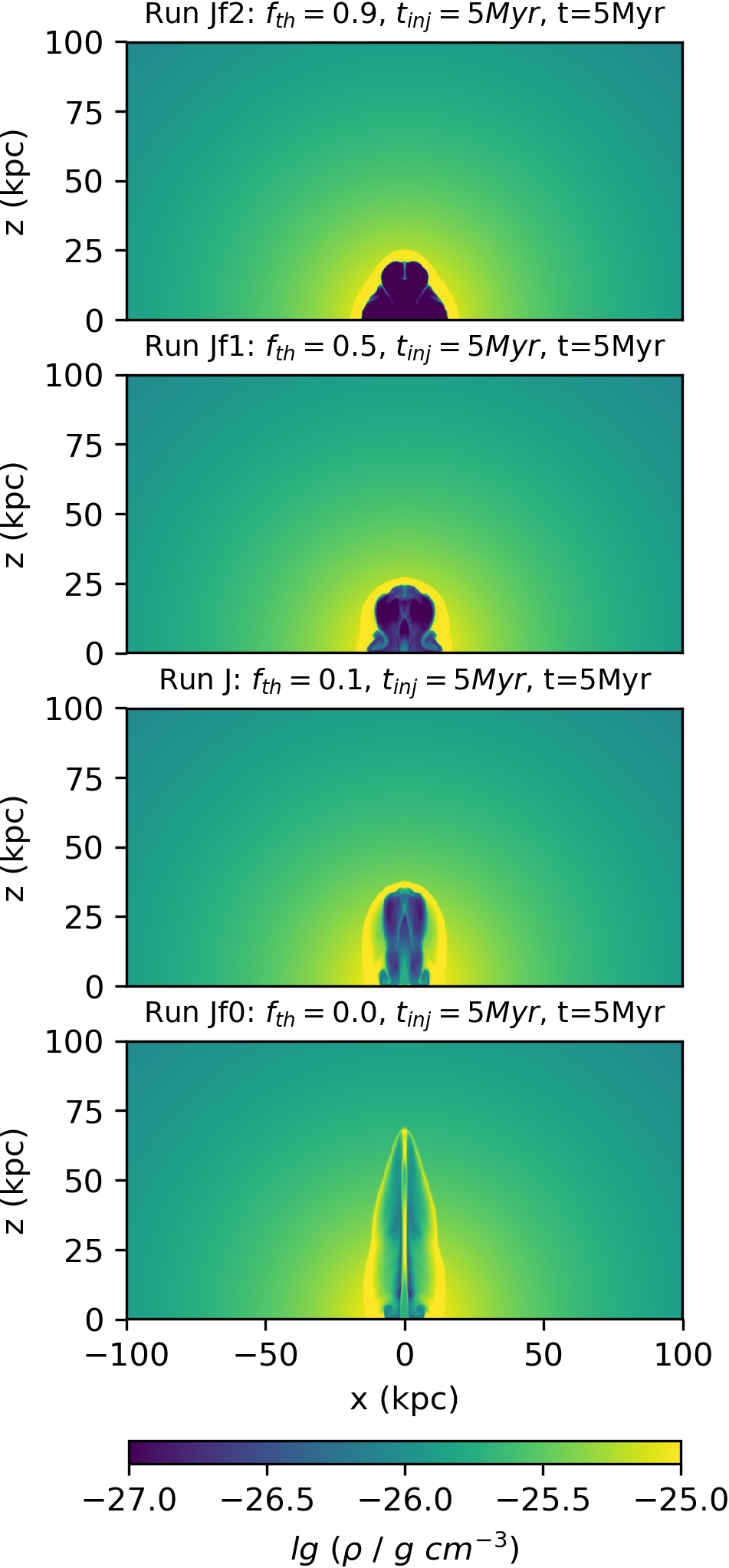}
\includegraphics[height=0.55\textheight]{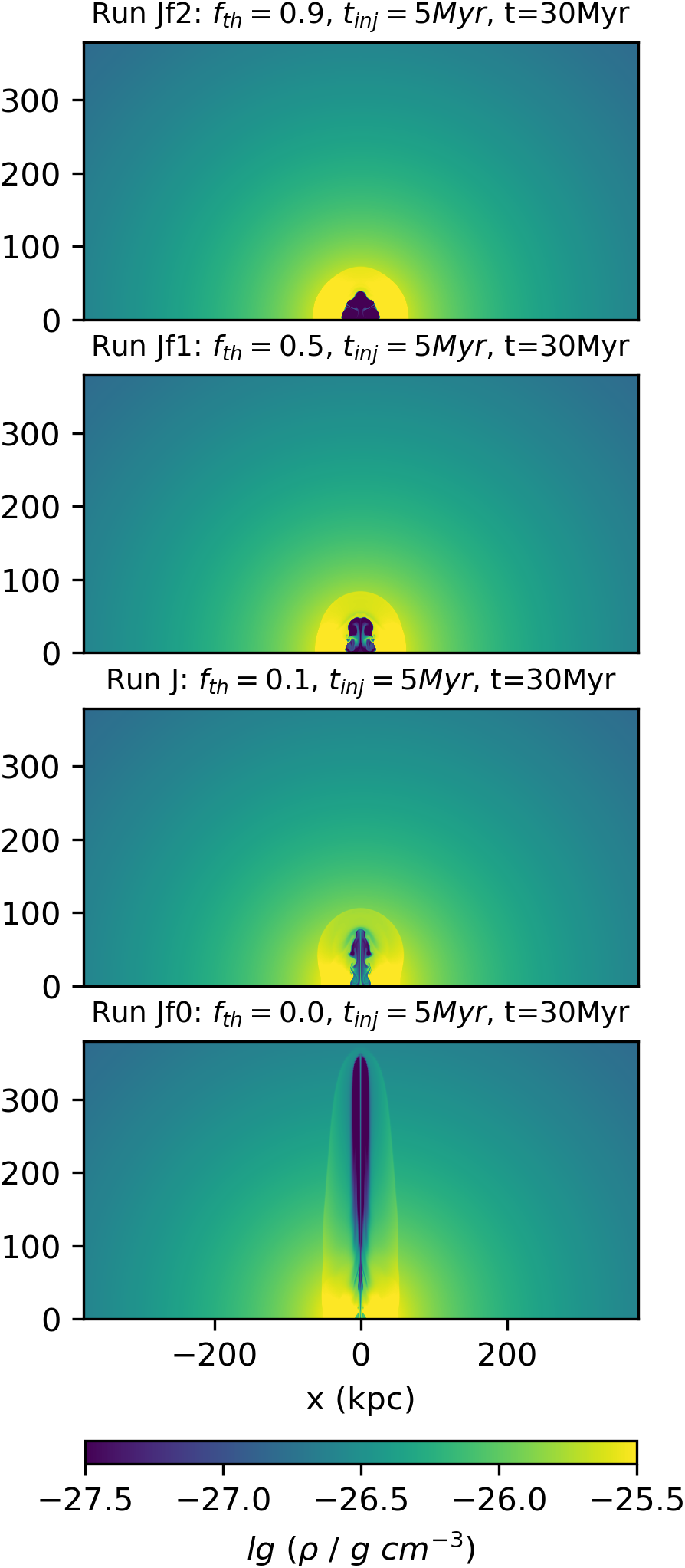}
\includegraphics[height=0.55\textheight]{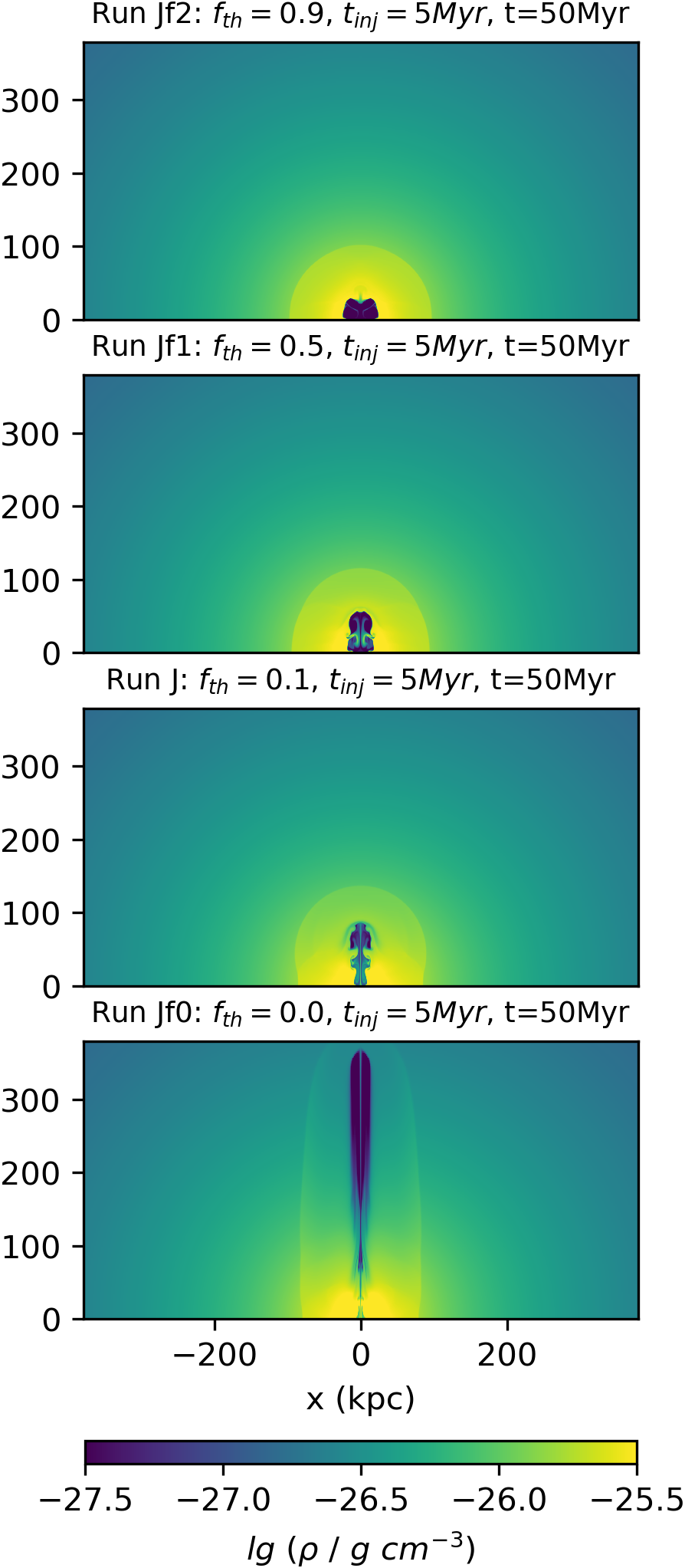}
}
\caption{Temporal evolution of AGN jet outbursts in a series of four strong outbursts with $P_{\rm inj}=10P_{\rm fb}$ ($t_{\rm inj}$=5Myr). From bottom to top, the four rows refer to four runs with thermal fraction $f_{\rm th}=0$, $0.1$, $0.5$, and $0.9$, respectively. From left to right, the three columns show the time evolution of the density distribution. Note that the panels in the first column shows the density distributions within the inner 100 kpc at $t=t_{\rm inj}$ when the jet injection just ends.}
\label{plot4}
\end{figure*}

In this subsection, we briefly describe the ejecta and shock morphologies in our jet outburst simulations with a focus on their dependences on thermal fraction $f_{\rm th}$ and the injection duration $t_{\rm inj}$ (see \citealt{guo15} and \citealt{guo18} for more relevant studies). The jet parameters and some results are listed in Table 2. The ejecta and shock morphologies are shown in Figure 4 and Figure 5 for typical powerful jets and mild jets respectively. 

In Figures 4 and 5, each row represents the outburst evolution in a specific simulation with a specific  thermal fraction $f_{\rm th}$, which is chosen to be $0$, $0.1$, $0.5$, and $0.9$, increasing from bottom to top. As is clearly seen, for a given outburst energy at a specific time, e.g., $t=50$ Myr, the travel distance of the jet ejecta along the jet direction decreases with $f_{\rm th}$ and increases with $t_{\rm inj}$. Both $f_{\rm th}$ and $t_{\rm inj}$ affect the thermal pressure ratio $p_{\rm inj} /p_{0}$ between the jet and the ambient medium at the jet base, as listed in Table 2. The travel distance of the jet ejecta is significantly affected by the value of $p_{\rm inj} /p_{0}$, and a larger pressure ratio induces stronger transverse expansion of the ejecta, which then receive stronger ram pressure and entrain more gas while traveling in the ambient ICM. Furthermore, for given values of $E_{\rm inj}$ and $f_{\rm th}$, mild outbursts with higher values of $t_{\rm inj}$ tend to produce larger ejecta bubbles than powerful outbursts with lower values of $t_{\rm inj}$. This is mainly due to stronger internal dissipation within the ejecta bubbles in the former case, as further shown in Section 4.4.     

Another result that we should pay attention to is the different shock structures in simulations with different jet powers. In the left column of Figure 4, the powerful jet outbursts with $P_{\rm inj}=10P_{\rm fb}$ are still in the early stage, clearly producing thin shocked shells in the ICM, and thus fall in the regime of the thin shell approximation investigated in Section 2.2. For the mild jet outbursts with $P_{\rm inj}=P_{\rm fb}$ shown in the left column of Figure 5, the thin shocked shell approximation may only be appropriate in front of the jet's working surface. For the kinetic-energy-dominated mild outbursts in the bottom two rows of Figure 5, one can see many substructures such as sound-wave ripples, weak shocks and even a second bow shock in run Jt1f0 at $t=100$ Myr, which can be more clearly seen in a much milder jet outburst shown in Figure A1 in Appendix A.

\begin{figure*}
\centering
\gridline{
\includegraphics[height=0.55\textheight]{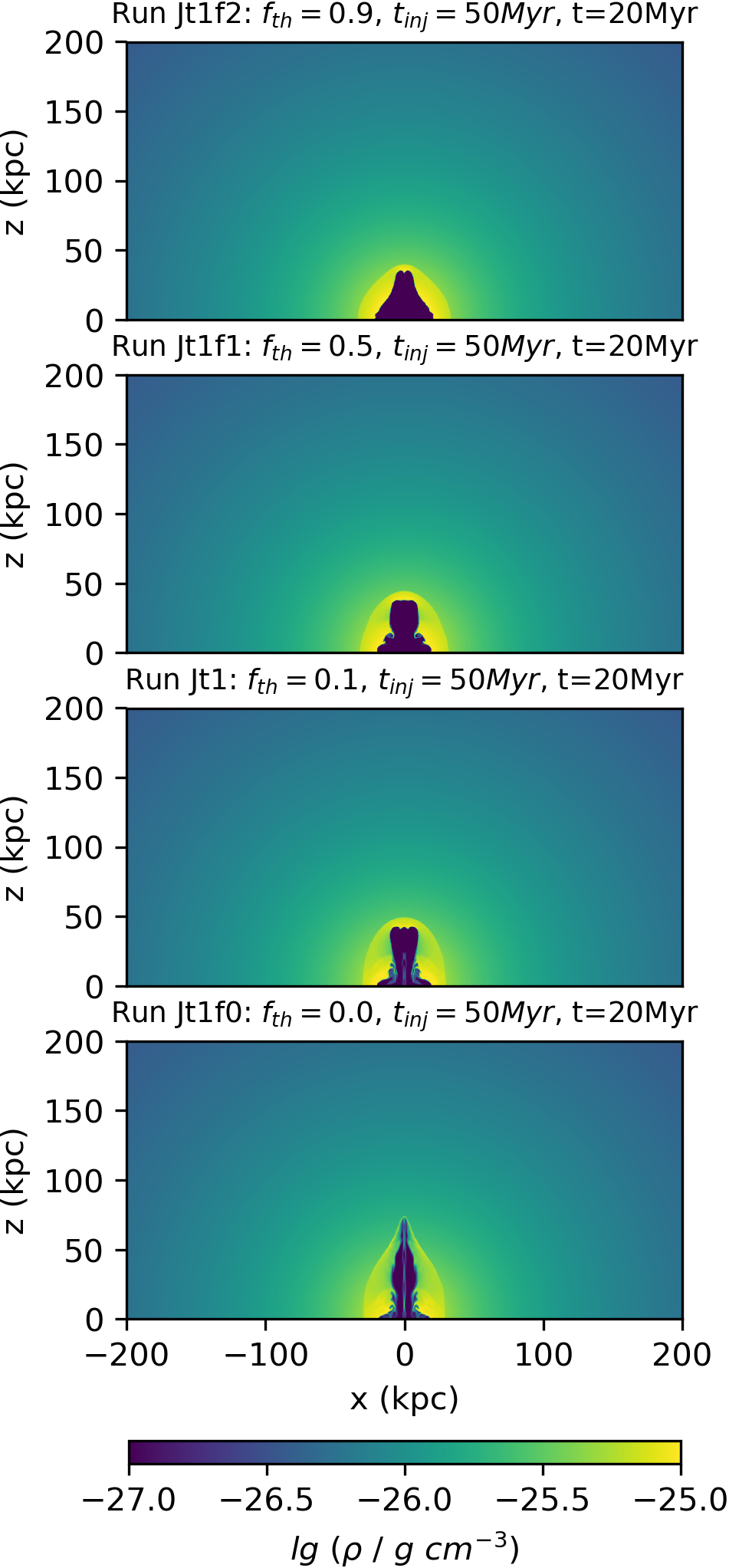}
\includegraphics[height=0.55\textheight]{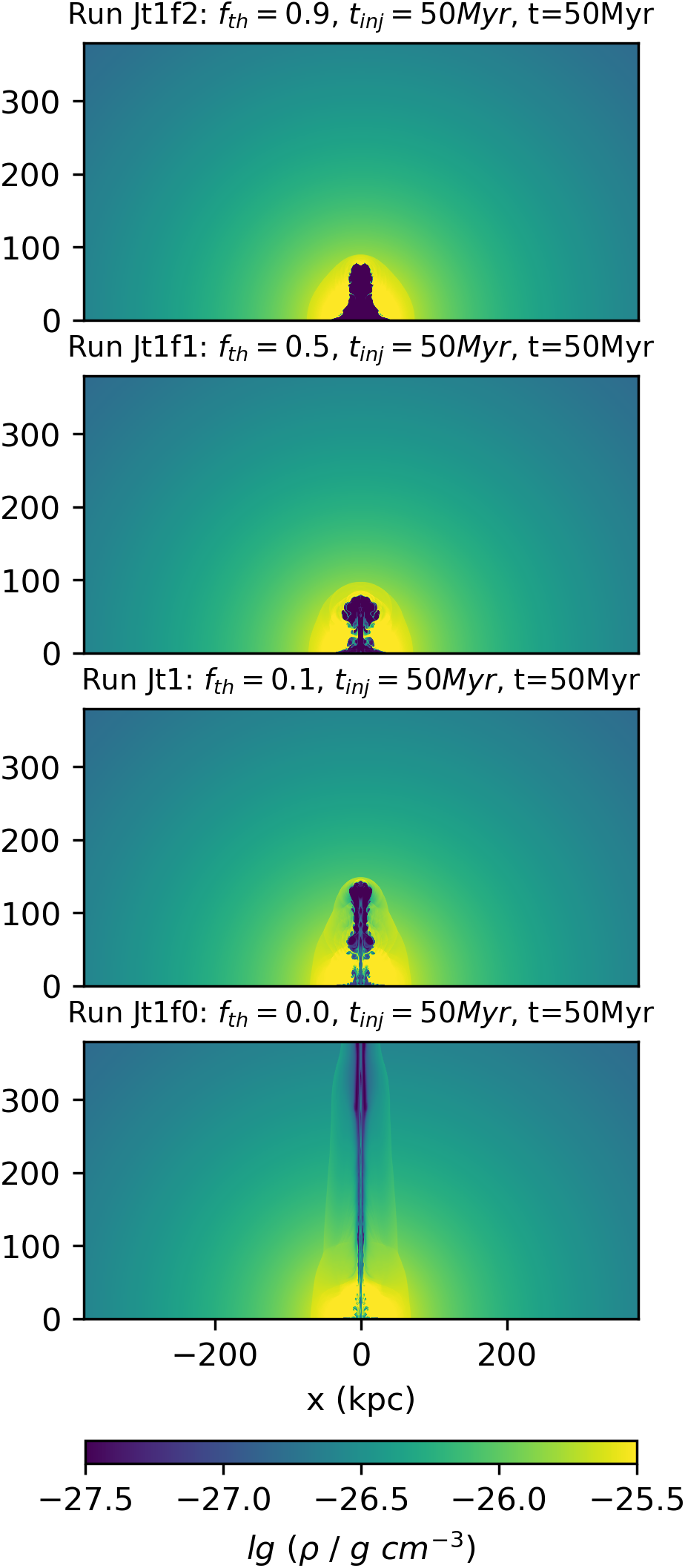}
\includegraphics[height=0.55\textheight]{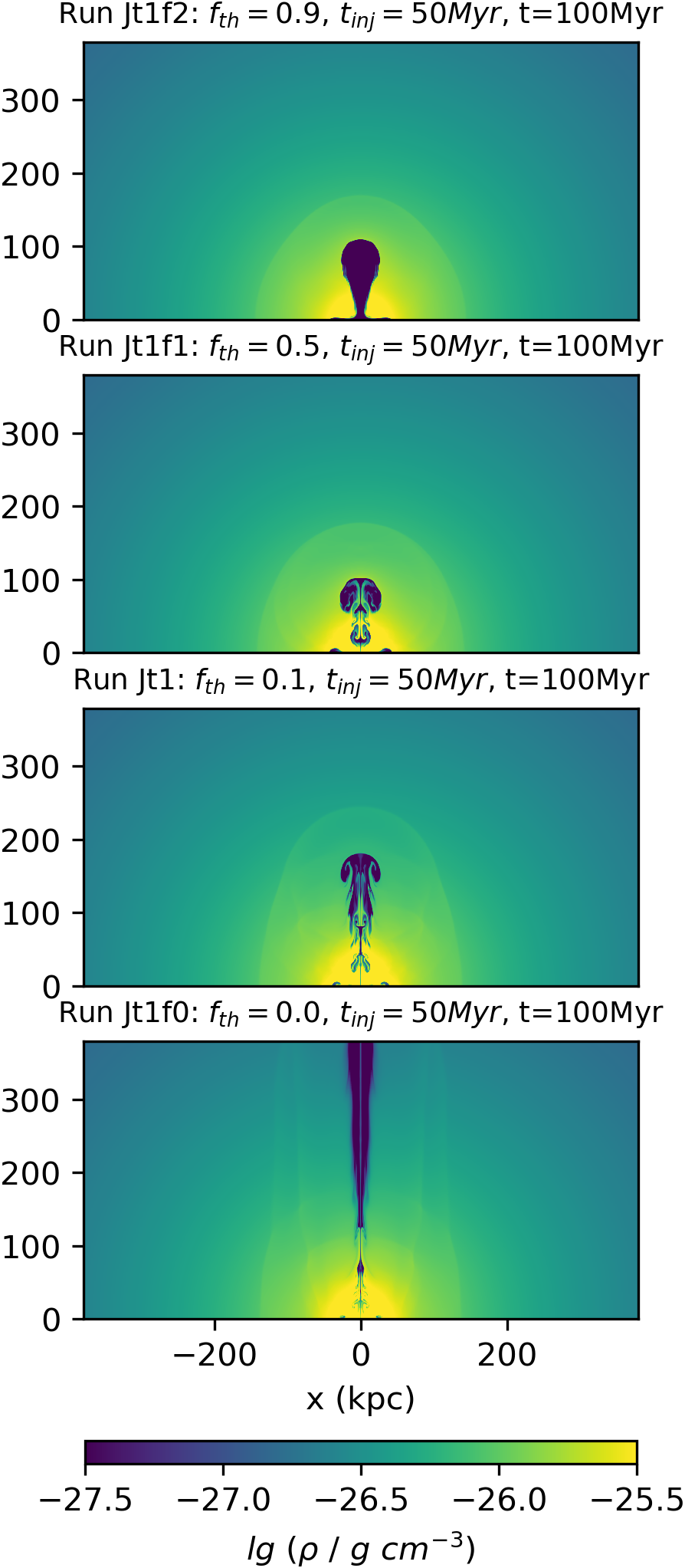}
}
\caption{Same as Figure 4, but for mild outbursts with  $P_{\rm inj}=P_{\rm fb}$ ($t_{\rm inj}=t_{\rm s}=50$ Myr). Note that the first column shows the gas density distributions within the inner 200 kpc at $t=20$ Myr, and the second column refers to the time $t=t_{\rm inj}$ Myr when the jet injection just ends.} 
\label{plot5}
\end{figure*}

\begin{figure*}
\centering
\gridline{
\includegraphics[width=0.258\textwidth]{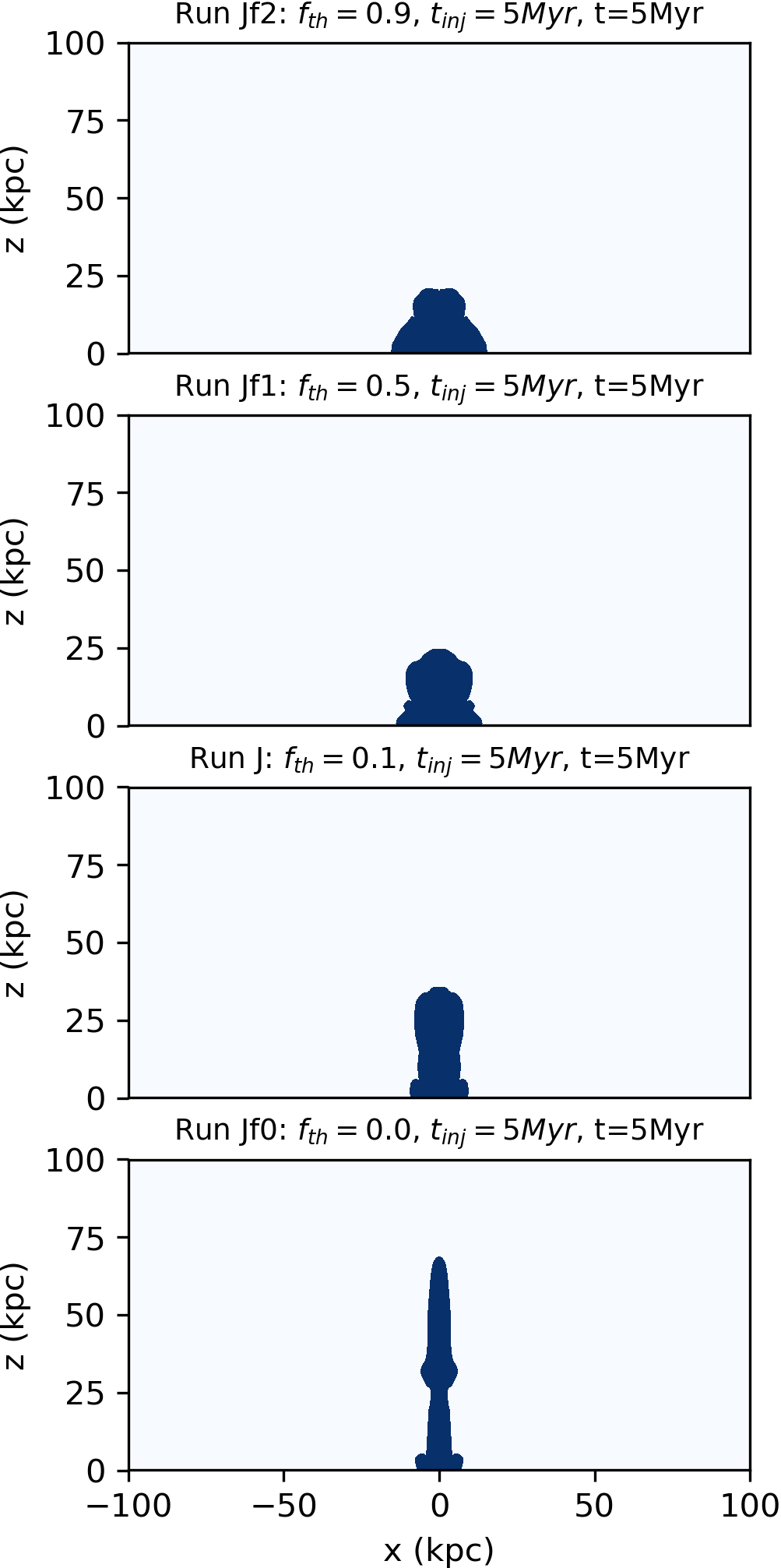}
\includegraphics[width=0.234\textwidth]{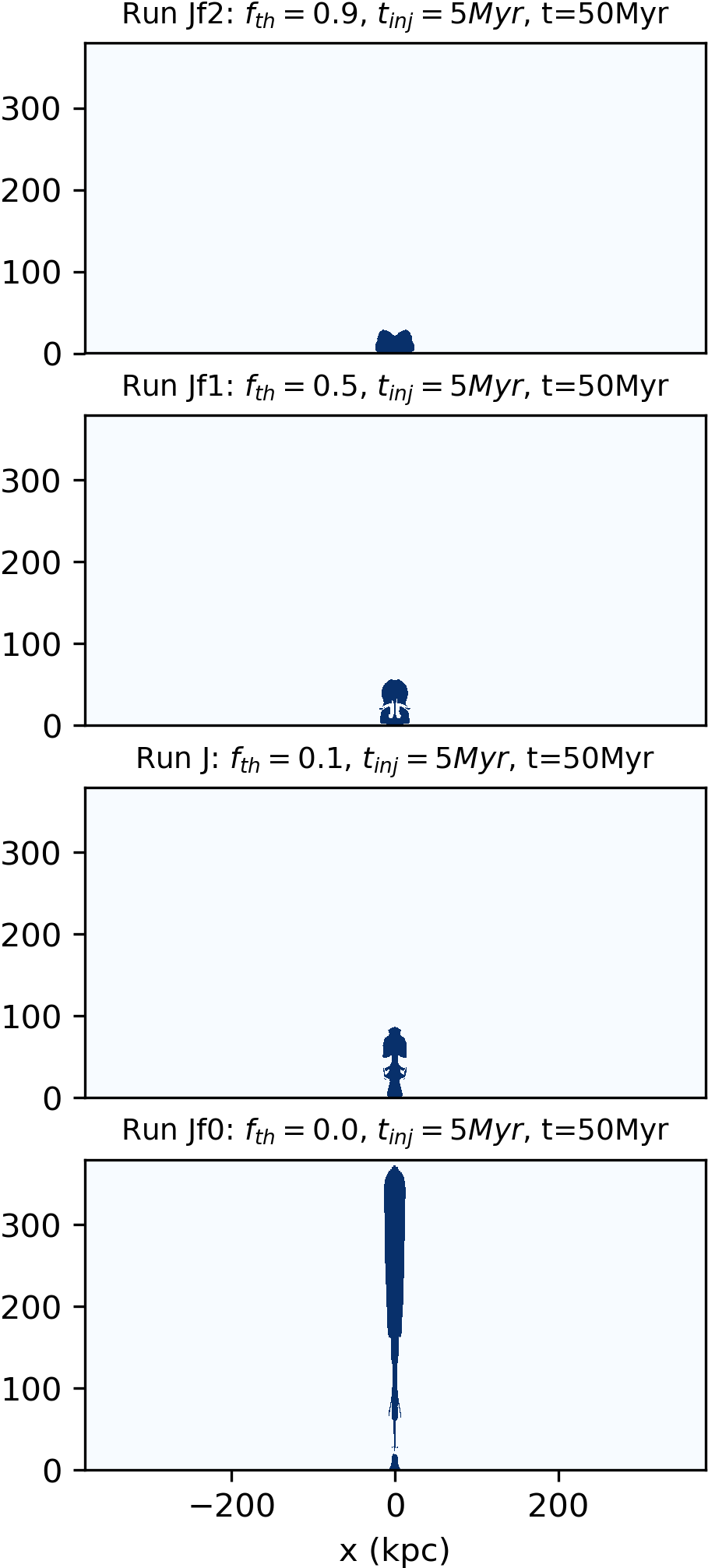}
\includegraphics[width=0.234\textwidth]{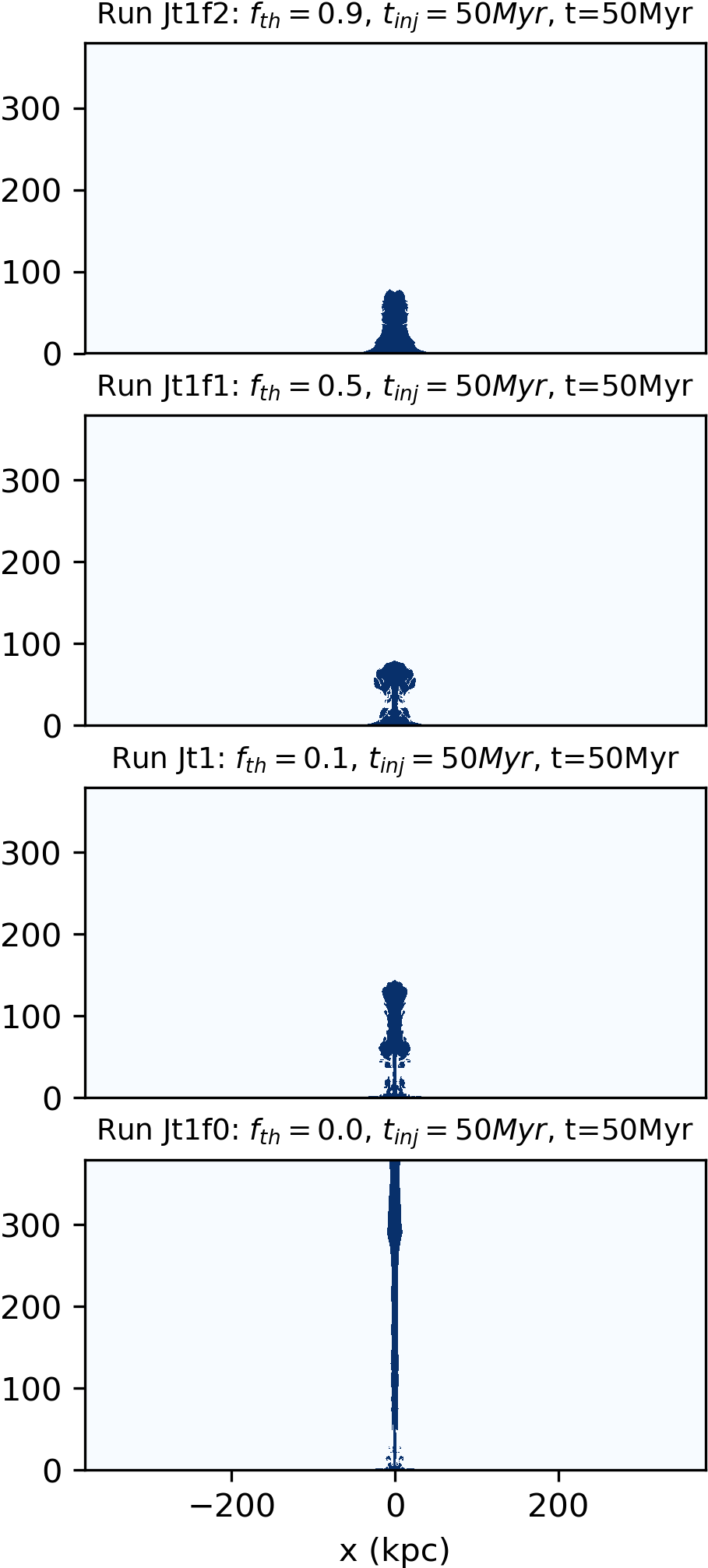}
\includegraphics[width=0.234\textwidth]{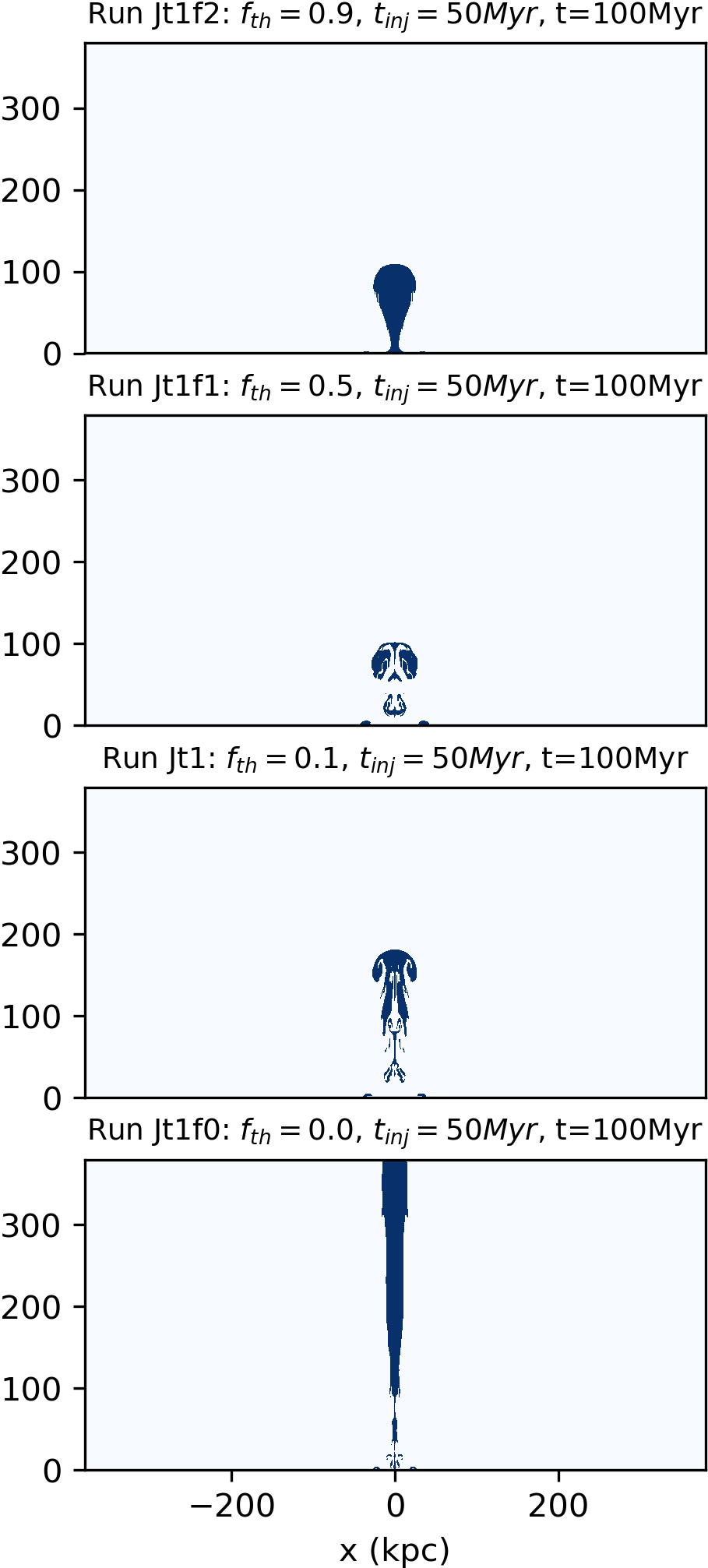}
}
\caption{Some representative ejecta bubbles identified with $M_{\rm bubble}/M_{\rm inj}<4$ using the scaler tracer method described in Section 4.3. These bubbles reproduce the corresponding low-density ejecta bubbles seen in Figure 4 and Figure 5 very well. }
\label{plot6}
\end{figure*}

\subsection{Energy Coupling Efficiency}
\label{section4.3}

To determine the temporal evolution of the energy coupling efficiency, we should identify the region of the jet ejecta bubbles on the fly in our jet simulations. To this end, we track the evolution of the ejecta bubble with a scalar tracer variable $\phi$ (\citealt{saxton01}; \citealt{duan18}) with an additional mass constraining method. The value of $\phi$ is constant along the trajectory of each gas element, and is set uniformly as $\phi=1$ in the ICM and $\phi=0$ in the jet at its base. Due to the unavoidable mixing between the jet ejecta and the ambient ICM, we identify the ejecta bubble as the region where $\phi < \phi_{0}$ and different values of $\phi_{0}$ ($0<\phi_{0}<1$) result in different masses of the identified ejecta bubble. Thus $\phi_{0}$ is a function of the mass of the identified bubble $\phi_{0}=\phi_{0}(M_{\rm bubble})$, and in this subsection, the ratio between $M_{\rm bubble}$ and the total injected jet mass $M_{\rm inj}$ is chosen to be 2 or 4. For these two choices of $M_{\rm bubble}$, the energy coupling efficiency in our simulations does not differ substantially, as clearly seen in the upper left panel of Figure 7 (see the difference between the red dotted and the red dashed lines corresponding to the total thermal energy within the ejecta bubble identified as $M_{\rm bubble}/M_{\rm inj}<2$ and $M_{\rm bubble}/M_{\rm inj}<4$ respectively). Some representative ejecta bubbles identified with $M_{\rm bubble}/M_{\rm inj}<4$ are shown in Figure 6, and they reproduce the corresponding low-density ejecta bubbles in Figure 4 and Figure 5 very well.  

One caveat that we note here is that as a typical finite-difference hydrodynamic code, our code does not guarantee energy conservation. While the total energy is conserved very well during most time of our simulations, we find that during the jet injection stage the total energy increase in the system is slightly lower than the total injected jet energy, a problem also mentioned in some recent studies (e.g., \citealt{bambic19}; \citealt{english19}). The difference increases with decreasing $f_{\rm th}$, and is typically less than $10\%$. To accurately calculate the energy partition and the energy coupling efficiency, we use the total energy increase in the system after the jet injection stage as the total injected jet energy.

To illustrate how the energy partition between the ejecta and the ICM evolves with time, we show the results of two representative simulations in the top panels of Figure 7. Run J represents a kinetic-energy-dominated powerful jet outburst, while run Jt1 represents a kinetic-energy-dominated mild jet outburst. The results of some additional runs are further presented in Figure A2 in Appendix A. One common feature in these simulations is that the total energy increase in the ICM (the black solid line) rises quickly during the jet injection, and then becomes relatively flat afterwards. In other words, the energy exchange between the jet ejecta and the ICM mainly occurs during and shortly after the jet injection, and the integrated energy coupling efficiency does not vary substantially at later times.

Table 2 lists the energy coupling efficiency $\eta_{\rm cp}$ evaluated at $t=t_{\rm inj}+t_{\rm s}$ in all our jet simulations, where $t_{\rm s}=50$ Myr and the ejecta bubble is identified with $M_{\rm bubble}/M_{\rm inj}<4$. One can directly see that $\eta_{\rm cp}\sim 0.7 - 0.9$ in all our simulations, scanning through a very large parameter space of $E_{\rm inj}$, $t_{\rm inj}$, $f_{\rm th}$ and $M_{\rm inj}$, which can also be seen in the left panel of Figure 8. This is very different from spherical outbursts presented in Section 3.2, where $\eta_{\rm cp}$ reaches the lower limit of $\eta_{\rm cp} \sim 0.4$ for slow isobaric outbursts. For jet outbursts, $\eta_{\rm cp}\sim 0.7 - 0.9$ is always very large for both powerful and weak outbursts. Most notably, in run Jt3 where the jet outburst is very weak with $t_{\rm inj}=300$ Myr and $P_{\rm inj}=P_{\rm fb}/6$, the value of $\eta_{\rm cp}$ is $0.86$, much higher than the expected value of  $\eta_{\rm cp} \sim 0.4$ for slow isobaric spherical outbursts. 

We argue that this difference is mainly caused by the non-spherical nature of jet outbursts, which produce hot spots at the jet's working surface. The hot spots dissipate the kinetic energy into thermal energy, and thus expand in the transverse direction, leading to significant backflows and enhancing the energy exchange between the jet ejecta and the ambient ICM. The non-spherical nature of jet outbursts also significantly enhances the generation of sound waves during the jet-ICM interaction, as shown recently by \citet{bambic19}.

\subsection{Outburst Evolution and Shock Heating}
\label{section4.4}

\begin{figure*}
\centering
\gridline{
\includegraphics[width=0.4\textwidth]{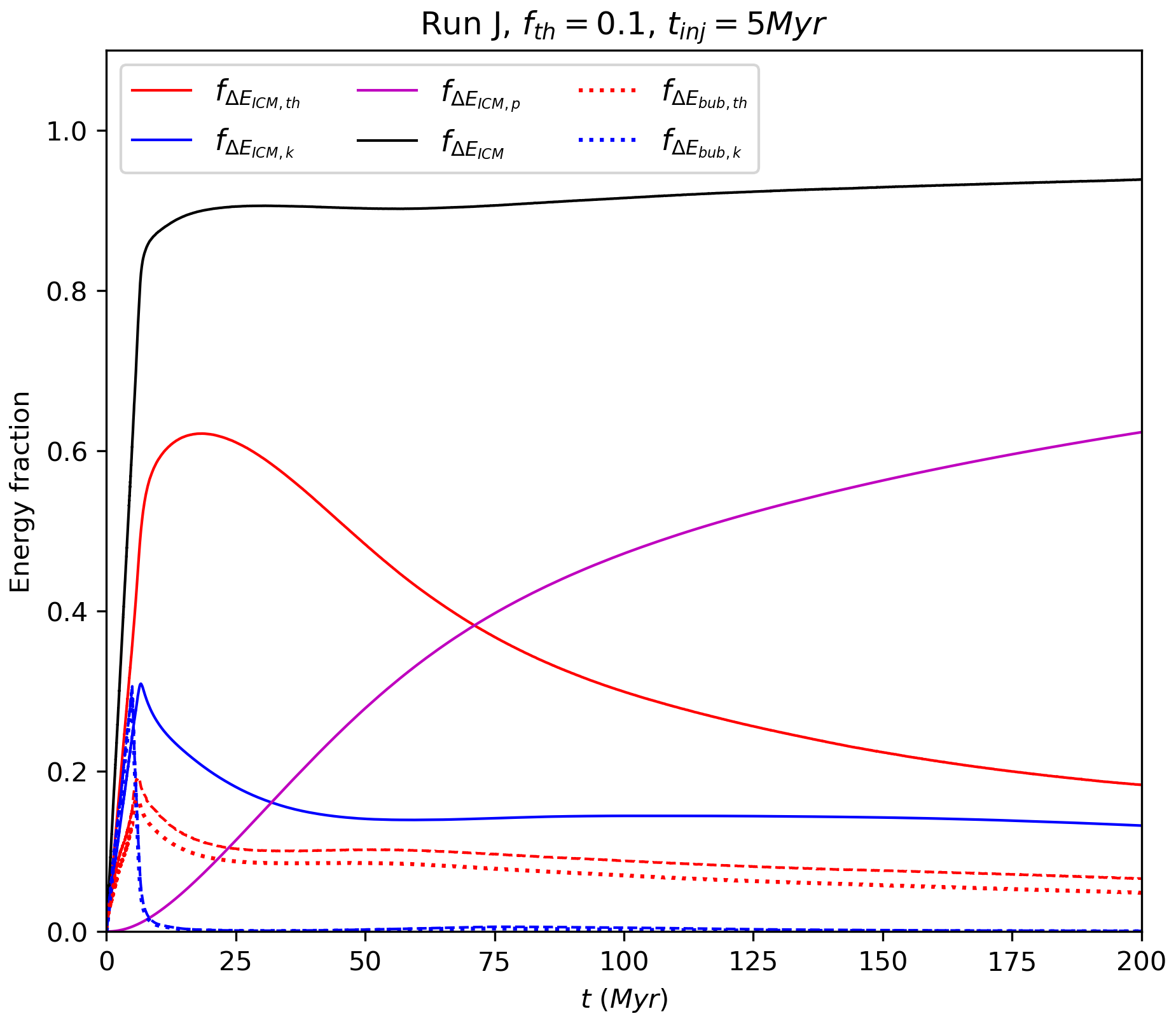}
\includegraphics[width=0.4\textwidth]{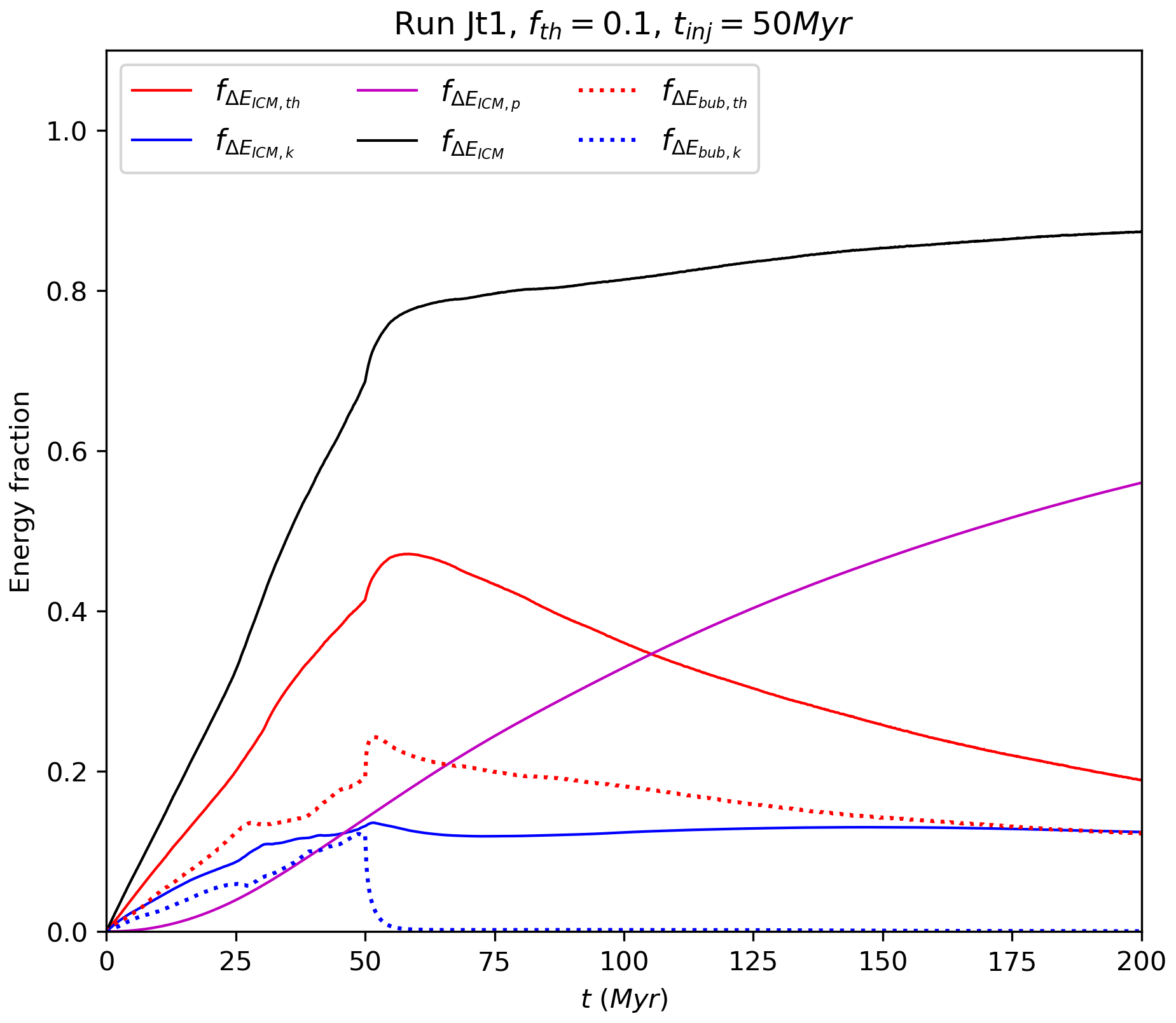}
}
\gridline{
\includegraphics[width=0.4\textwidth]{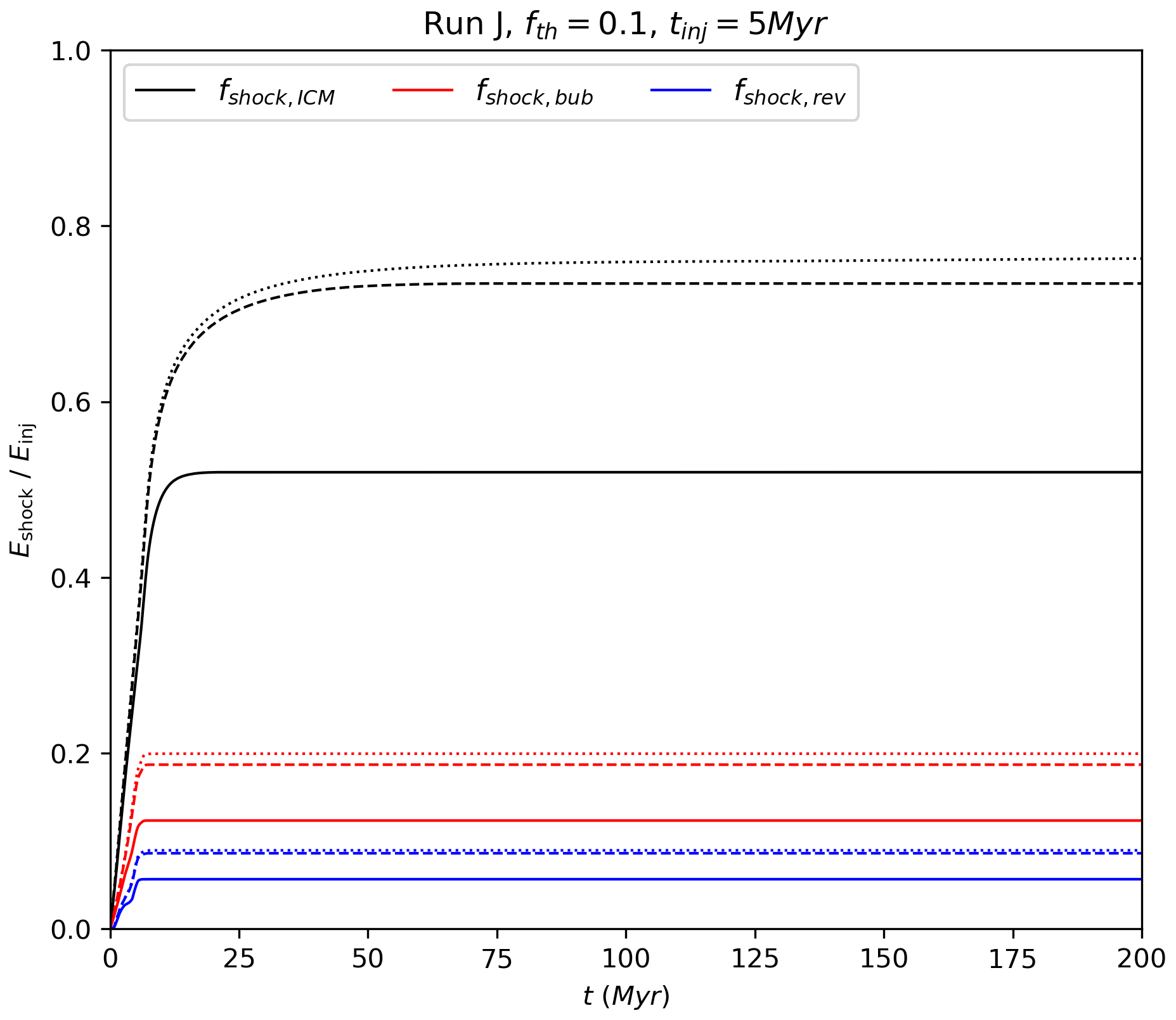}
\includegraphics[width=0.4\textwidth]{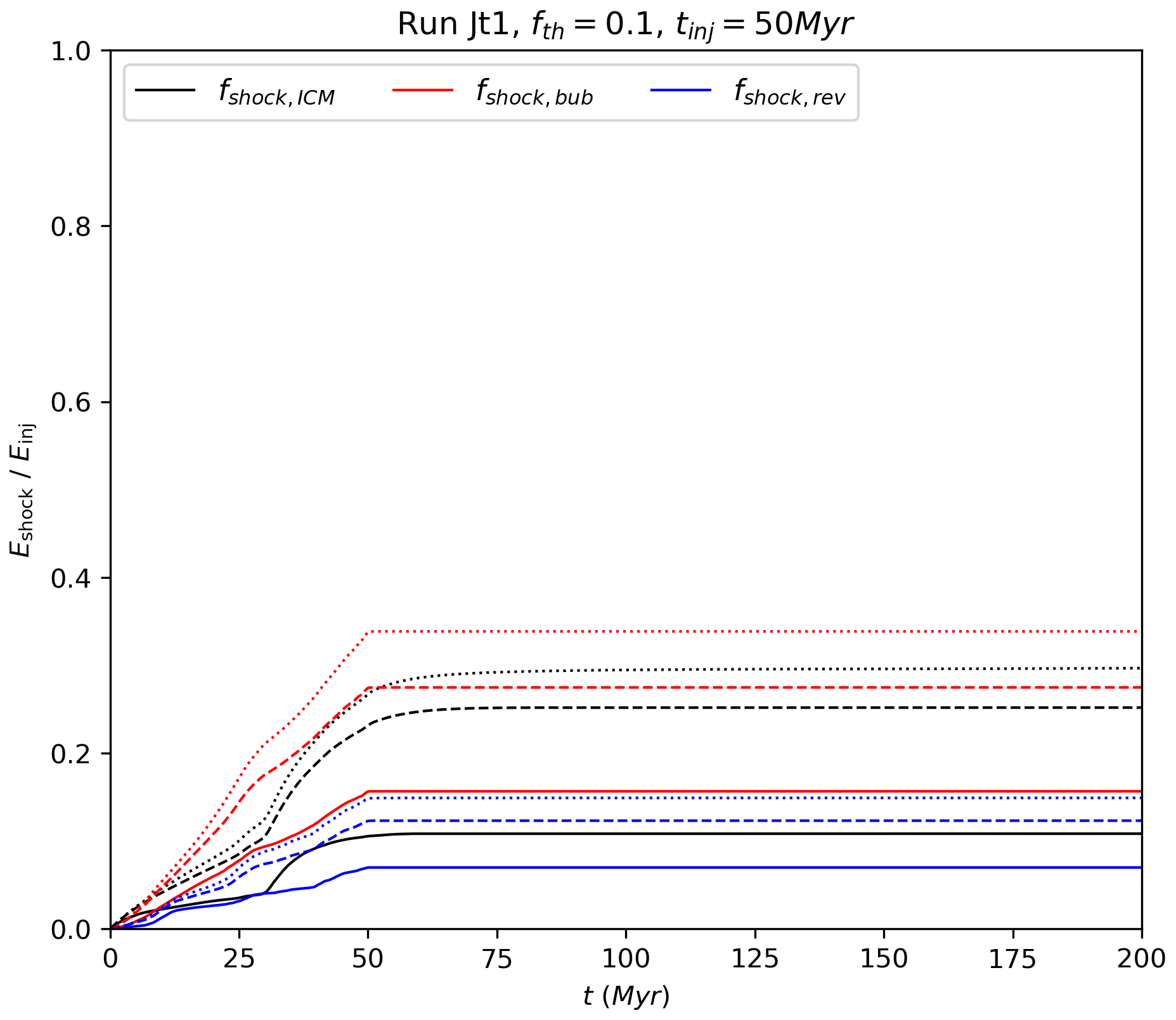}
}
\caption{Temporal evolution of energy partition (top panels) and time-integrated energy dissipated by shocks (bottom panels). The energy fractions in the top panels are normalized by the total injected energy $E_{\rm inj}$ during the whole jet outburst. The subscript `ICM' refers to the ICM (beyond the ejecta bubble), while the subscript `bub' refers to the ejecta bubble. The red dotted and dashed lines in the top-left panel correspond to the total thermal energy within the ejecta bubble identified as $M_{\rm bubble}/M_{\rm inj}<2$ and $M_{\rm bubble}/M_{\rm inj}<4$ respectively, while the former criterion is adopted to identify the ejecta bubble for all the other lines in the top panels. In the bottom panels, shock dissipation is evaluated in all active zones with pressure jumps across two adjacent zones $p_{2} / p_{1}>1.6$ (solid), $1.1$ (dashed), and $1.01$ (dotted). The line for $p_{2} / p_{1}>1.001$ roughly coincides with the dotted line and is not shown here. The blue lines refer to energy dissipation by the reverse shock, which mainly occurs within the ejecta bubble. Note that the shock front is typically resolved by several zones in our simulations. }
\label{plot7}
\end{figure*}

\begin{figure*}
\centering
\includegraphics[height=0.28\textheight]{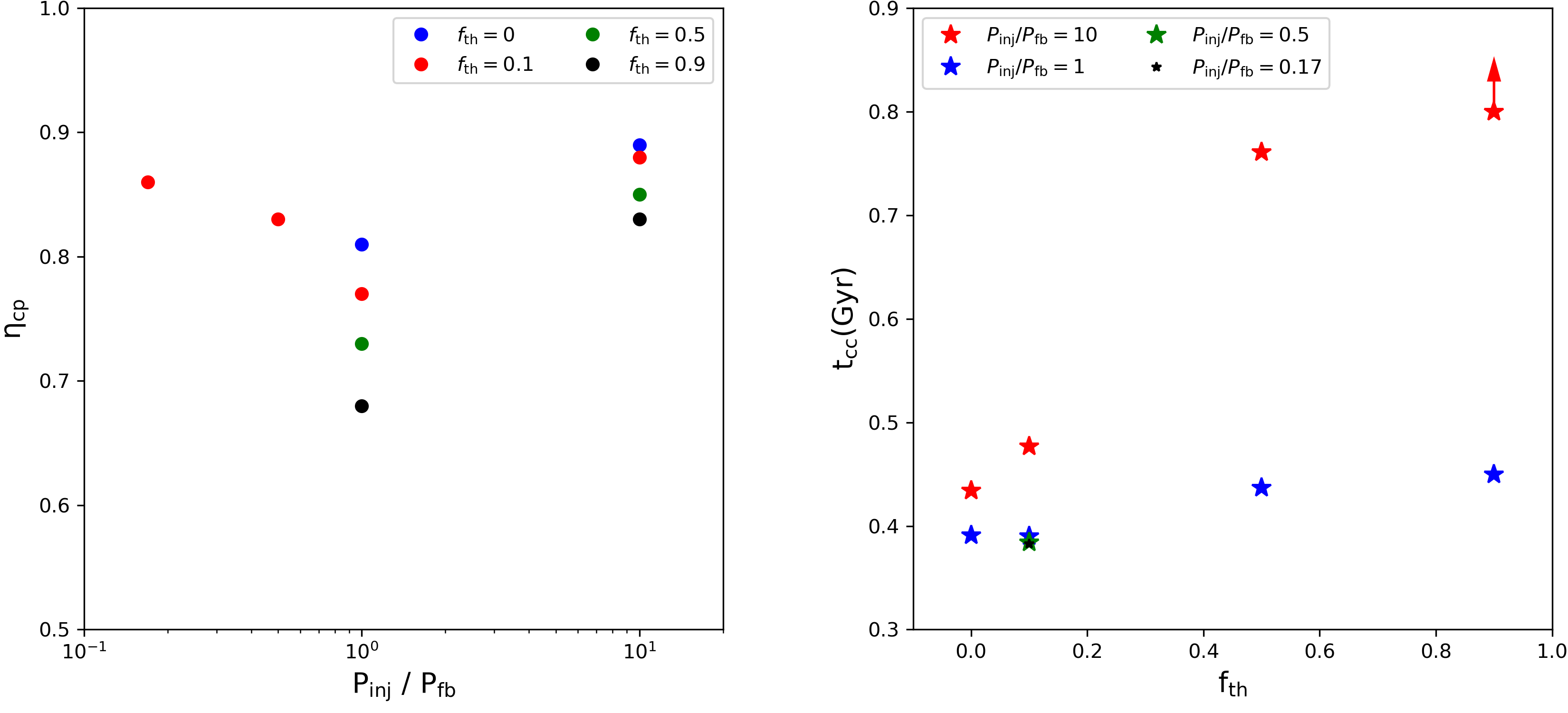}
\caption{ Dependence of the energy coupling efficiency $\eta_{\rm cp}$ (left) and the start time of the central cooling catastrophe $t_{\rm cc}$ (right) on the jet power and thermal fraction $f_{\rm th}$. The data points in this Figure refer to the results of the first 10 simulations listed in Table 2, which all have the same total jet energy $E_{\rm inj}$ and Mach number $M_{\rm inj}$. $P_{\rm inj}/P_{\rm fb}$ is the ratio of the jet power to the characteristic jet power $P_{\rm fb}$ as defined in Section 4.1. As indicated in the legend, the color of each data point in the left panel refers to thermal fraction $f_{\rm th}$ adopted in the corresponding simulation, while that in the right panel stands for $P_{\rm inj}/P_{\rm fb}$. The left two data points in the left panel indicates that $\eta_{\rm cp}$ is larger than $0.8$ even when the jet power is substantially lower than $P_{\rm fb}$.
}
\label{plot8}
\end{figure*}

In this subsection, we further investigate the evolution of the outburst energy partition and the efficiency of shock heating in our jet simulations. As in \citet{guo18}, shock dissipation in our simulations is implemented with a von Neumann-Richtmyer artificial viscosity (same as the ZEUS code; \citealt{zeus2d92}; \citealt{li17}). The shock fronts are identified by pressure jumps across adjacent grids, and our calculations include both forward and reverse shocks. The temporal evolution of the time-integrated energy dissipated by shocks in our simulations are shown in the bottom panels of Figures 7 and A2. As shown in Figures 7 and A2, both the ambient ICM and the jet ejecta are heated by shocks. The internal dissipation in mild jet outbursts is usually stronger than in powerful outbursts, which may explain why, with the same jet energy, the ejecta bubbles in the former case are typically larger. The temporal evolutions of profiles of gas pressure, density and velocity along the $z$ and $R$ directions in three representative runs are illustrated in Figures A3 and A4 in the Appendix, respectively. 

According to the temporal evolution of the energy partition, the outburst evolution may be divided into three stages as follows. We pay particular attention to the differences between powerful and mild outbursts. 

(a) Injection stage. During this stage, the jet is actively blowing an ejecta bubble, transferring its energy to the ambient ICM by increasing the ICM's thermal and kinetic energies. As the ambient ICM is pushed outward, the thermal and kinetic energies of the ICM are continuously converted to its gravitational energy, which is especially significant for mild outbursts with a long jet duration, as clearly seen in Figure 7. For powerful jet outbursts, a large fraction of the ICM's kinetic energy is dissipated at the forward shock, replenishing the entropy lost through radiative cooling, and the total shock-dissipated energy in the ICM reaches about $70\%$ of the injected jet energy $E_{\rm inj}$ in run J. On the other hand, for mild jet outbursts such as in run Jt1, the shock-dissipated energy within the ICM is subdominant, accounting for only $\lesssim 30\%$ of $E_{\rm inj}$, although the total jet energy transferred to the ICM reaches about $70\%$ of $E_{\rm inj}$ at $t=t_{\rm inj}$. During the jet evolution, backflows emanating from the hotspots interact dynamically with the ambient ICM, continuously and efficiently transferring the ejecta energy to the ICM. While a large fraction of this energy is eventually dissipated by relatively strong shocks driven by powerful outbursts, mild outbursts induce shocks with low Mach numbers which are inefficient in dissipating the kinetic energy. 

(b) Braking stage. As the jet injection ends, the ejecta bubble experiences abrupt braking, which is manifested as the abrupt decreasing of its internal kinetic energy shown in the top panels of Figure 7. This feature can also be seen through the significant decrease of the z-component velocity $v_{\rm z}$ after $t=t_{\rm inj}$ shown in the bottom panels of Figure A3 in Appendix A. During this brief braking stage, the ejecta's kinetic energy is mainly transferred to its thermal energy and the ambient ICM's thermal energy, as shown in Figures 7 and A2. For strong jet outbursts, the braking occurs as the shock-heated ambient gas is still mainly in a thin shell beyond the ejecta bubble, as illustrated in the left column of Figure 4, while for mild outbursts shown in the middle column of Figure 5, the thin-shell approximation may only be applicable near the jet's working surface. During and shortly after the braking stage, the ejecta bubble continues to expand, losing its thermal energy to the ambient ICM, which is especially significant for thermal-energy-dominated jet outbursts, as clearly shown in the middle and right columns of Figure A2 in Appendix A.

(c) Rising stage. At this stage, the ejecta bubble detaches from the jet base and rises towards large radii buoyantly. The main feature shown in Figures 7 and A2 during this stage is that the total energy in the ICM (the black solid line) changes very slowly. In other words, the energy exchange between the ejecta bubble and the ICM mainly occurs during the injection and braking stages, and the $pdV$ work done by the ejecta during the rising stage is unimportant for AGN feedback energetics. We also note that during the rising stage, the ICM continues to expand, resulting in the continuous conversion of its thermal energy to the gravitational energy.

\subsection{Impact on the Cooling Catastrophe}
\label{section4.5}

In all our jet simulations investigated above, radiative cooling is not included. To investigate the impact of AGN outbursts on the development of cooling flows, we reran all the jet outburst simulations with radiative cooling included. We adopt the same setup of radiative cooling as in \citet{guo18}. We find that in all our jet simulations, the energy coupling efficiency $\eta_{\rm cp}$ changes very little when radiative cooling is included.

Without heating sources, radiative cooling leads to a gradual decrease in the ICM temperature and a subsequent cooling catastrophe in the central region at a later time (e.g., \citealt{guo14}; \citealt{guo18}). As the cooling catastrophe occurs, the gas temperature in the central region drops very quickly and cold gas quickly accumulates. In this paper, we define the start time of the central cooling catastrophe $t_{\rm cc}$ as the time when the total mass of cold gas within the central $\rm 1$ kpc with temperature below $5 \times 10^{5}$ K reaches $10^{6} \rm M_{\sun}$. As shown in \citet{guo18}, without AGN outbursts, the central cooling catastrophe in our modeled ICM is expected to develop at $t_{\rm cc} = 238$ Myr. With an episode of AGN outburst at the beginning of each simulation, the development of the central cooling catastrophe is usually delayed and thus $t_{\rm cc}$ becomes larger.

The values of $t_{\rm cc}$ in our simulations are listed in the rightmost column of Table 2. As seen in Table 2, in most simulations except run JM1 where the injected gas density is very high, the cooling catastrophe is typically delayed for more than $100$ Myr. For the same injected energy $E_{\rm inj}$, the cooling catastrophe tends to be delayed for a longer duration when the jet power $P_{\rm inj}$ or thermal fraction $f_{\rm th}$ is higher. While the energy coupling efficiency $\eta_{\rm cp}$ typically decreases slowly with increasing $f_{\rm th}$, $t_{\rm cc}$ increases with $f_{\rm th}$, most notably in powerful outbursts as clearly seen in Figure 8. With similar values of $\eta_{\rm cp}\sim 0.7 - 0.9$, kinetic-energy-dominated jets travel to larger distances, depositing less energy within the cool core than thermal-energy-dominated jets \citep{guo16}. Furthermore, as seen in Figures 4 and 5, the ejecta bubbles produced by thermal-energy-dominated jets stay near the cluster center for much longer time, suppressing the gas accumulation at the cluster center and thus delaying the central cooling catastrophe. Figure 8 clearly indicates that powerful thermal-energy-dominated jets are most effective in delaying the onset of the central cooling catastrophe. Thermal-energy-dominated jets are internally-subsonic jets \citep{guo16}, which may be alternatively dominated by cosmic ray energy (\citealt{guo11}; \citealt{ruszkowski17}; \citealt{yang19}).

\section{Summary and brief discussions}
\label{section5}

Using both analytical and numerical methods, we present a systematic study on the energy coupling efficiency $\eta_{\rm cp}$ of AGN outbursts in the ICM. To get physical intuition,we first investigate spherical outbursts in a uniform medium. We estimate the values of $\eta_{\rm cp}$ in two extreme situations, including weak isobaric outbursts with $\eta_{\rm cp}\sim 0.4$ and powerful point outbursts with $\eta_{\rm cp}\gtrsim 0.8$ going through the classic Sedov-Taylor phase. We then investigate the energy coupling efficiency of spherical outbursts in a uniform medium with a suite of hydrodynamic simulations and perform a large parameter study over the total energy $E_{\rm inj}$, outburst duration $t_{\rm inj}$, thermal fraction $f_{\rm th}$, and Mach number $M$ of the outbursts. At last, we investigate AGN jet outbursts in a realistic ICM with a series of hydrodynamic simulations and perform a parameter study over the same large parameter space as for spherical outbursts. Our main conclusions are summarized as follows.

(i) Our spherical outburst simulations confirm that the energy coupling efficiency increases from $\sim 0.4$ for a weak outburst to $\gtrsim 0.8$ for a very powerful outburst. For any given outburst energy $E_{\rm inj}$, we identify two characteristic outburst powers $P_{\rm fb}$ and $P_{\rm tr}\sim 5P_{\rm fb}$ that roughly separate weak and powerful outbursts. $P_{\rm fb}$ is determined by the sound crossing time across the ambient medium region significantly affected by the outburst energy $E_{\rm inj}$. Outbursts with $P_{\rm inj} \ll P_{\rm fb}$ can be regarded as weak outbursts in the slow-piston limit with $\eta_{\rm cp}\sim 0.4$. For outbursts with $P_{\rm inj} > P_{\rm tr}$, the postshock gas right behind the induced forward shock remains supersonic with respect to the ambient medium during the whole outburst, and these outbursts can be regarded as powerful outbursts in the thin-shell approximation with $\eta_{\rm cp}\gtrsim 0.8$.

(ii) Jet outbursts are intrinsically different from spherical outbursts. Our jet simulations in a realistic ICM indicate that $\eta_{\rm cp}$ is typically around $0.7-0.9$ for both powerful and weak jet outbursts. This is caused by the non-spherical nature of jet outbursts, which produce significant backflows emanating from the hotspots, enhancing the energy exchange between the jet ejecta and the ambient ICM. From this result, one may estimate the jet outburst energy in X-ray observations of galaxy clusters according to 
\begin{eqnarray}  
E_{\rm jet}  \approx \frac{p_{\rm cav}V_{\rm cav}}{(1-\eta_{\rm cp})(\gamma_{\rm cav}-1)} ,
\end{eqnarray}
where $p_{\rm cav}$ and $V_{\rm cav}$ are the observed pressure and volume of X-ray cavities, respectively, and $\gamma_{\rm cav}$ is the effective adiabatic index of the plasma in X-ray cavities. The value of $\gamma_{\rm cav}$ depends on the dominant energy content within the cavities, and is still in debate (\citealt{boehringer93}; \citealt{croston14}; \citealt{blandford19}). If taking $\gamma_{\rm cav}=4/3$, the outburst energy can be estimated as $E_{\rm jet} \approx 10 - 30 p_{\rm cav}V_{\rm cav}$ for $\eta_{\rm cp} = 0.7 - 0.9$. If taking $\gamma_{\rm cav}=5/3$, $E_{\rm jet} \approx 5 - 15 p_{\rm cav}V_{\rm cav}$ for $\eta_{\rm cp} = 0.7 - 0.9$. We note that plasma radiation can also take part of the jet energy away, but it is usually considered to be unimportant in outburst energetics on the cluster scale (\citealt{birzan04}; \citealt{birzan08}; \citealt{cavagnolo10}; \citealt{osullivan11}).

(iii) The temporal evolution of AGN jet outbursts in the ICM may be divided into three stages: the injection, braking, and rising stages. The energy exchange between the ejecta bubble and the ICM mainly occurs during the first two stages, during which the ejecta energy is continuously transferred to the ambient ICM by increasing the ICM's thermal and kinetic energies. AGN outbursts induce the cool core expansion (\citealt{guo18}), during which the ICM's thermal and kinetic energies are gradually converted to its gravitational energy. 

(iv) For powerful jet outbursts, a large fraction of the acquired ICM kinetic energy is dissipated at the forward shock during the injection and braking stages, replenishing the entropy lost through radiative cooling. The total shock-dissipated energy in the ICM typically reaches about $70-80 \%$ and $50 \%$ of $E_{\rm inj}$ for kinetic-energy-dominated and thermal-energy-dominated jet outbursts, respectively (see Figures 7 and A2). On the other hand, for mild jet outbursts, the shock-dissipated energy within the ICM is subdominant, typically accounting for $\lesssim 30\%$ of $E_{\rm inj}$, although the total jet energy transferred to the ICM reaches about $70-80 \%$ of $E_{\rm inj}$. The efficiency of shock dissipation depends sensitively on its Mach number $M$ at $1<M<2$, and mild outbursts induce shocks with low Mach numbers which are inefficient in dissipating the kinetic energy. 

(v) While the energy coupling efficiency of powerful and weak outbursts is similar in the ICM and slightly decreases with $f_{\rm th}$ (see Table 2), powerful thermal-energy-dominated jets are most effective in delaying the onset of the central cooling catastrophe. For the same outburst energy, kinetic-energy-dominated jets travel to larger distances, depositing less energy within the cool core than thermal-energy-dominated jets. The ejecta bubbles produced by thermal-energy-dominated jets stay near the cluster center for much longer time, suppressing the gas accumulation at the cluster center and also contributing to the delay of the central cooling catastrophe.  

Both powerful and mild jet outbursts are efficient in transferring energy to the ambient ICM, but to solve the cooling flow problem, the transferred energy should be quickly transported to the whole cluster cool core with a typical radial size of $\sim 100 - 200$ kpc. The transported energy should also be efficiently dissipated locally across the cool core, replenishing the entropy lost through radiative cooling. Although extensively studied, the detailed mechanisms that transport and dissipate the outburst energy across the whole cool core remain to be elucidated by future studies.


\acknowledgements 

XD thanks for the discussions with Bocheng Zhu, Ruiyu Zhang and Shaokun Xie. FG thanks Eugene Churazov for very fruitful discussions. We thank the anonymous referee for a constructive report. This work was supported in part by the National Natural Science Foundation of China (No. 11873072 and 11633006), the Natural Science Foundation of Shanghai (No. 18ZR1447100), and the Chinese Academy of Sciences through the Key Research Program of Frontier Sciences (No. QYZDB-SSW-SYS033 and QYZDJ-SSW-SYS008). The simulations presented in this work were performed using the high performance computing resources in the Core Facility for Advanced Research Computing at Shanghai Astronomical Observatory.

\bibliography{ms} 

\appendix
\section{Additional Figures on AGN Jet Outbursts}

In this appendix, we present four additional figures (Figures A1 - A4) from our jet outburst simulations listed in Table 2. These figures show more details on AGN jet outbursts in our simulations, and are used in our discussions in Section 4.

\renewcommand\thefigure{\Alph{section}\arabic{figure}}
\setcounter{figure}{0} 

\begin{figure*}[h!]
\centering
\gridline{
\includegraphics[width=0.6\textwidth]{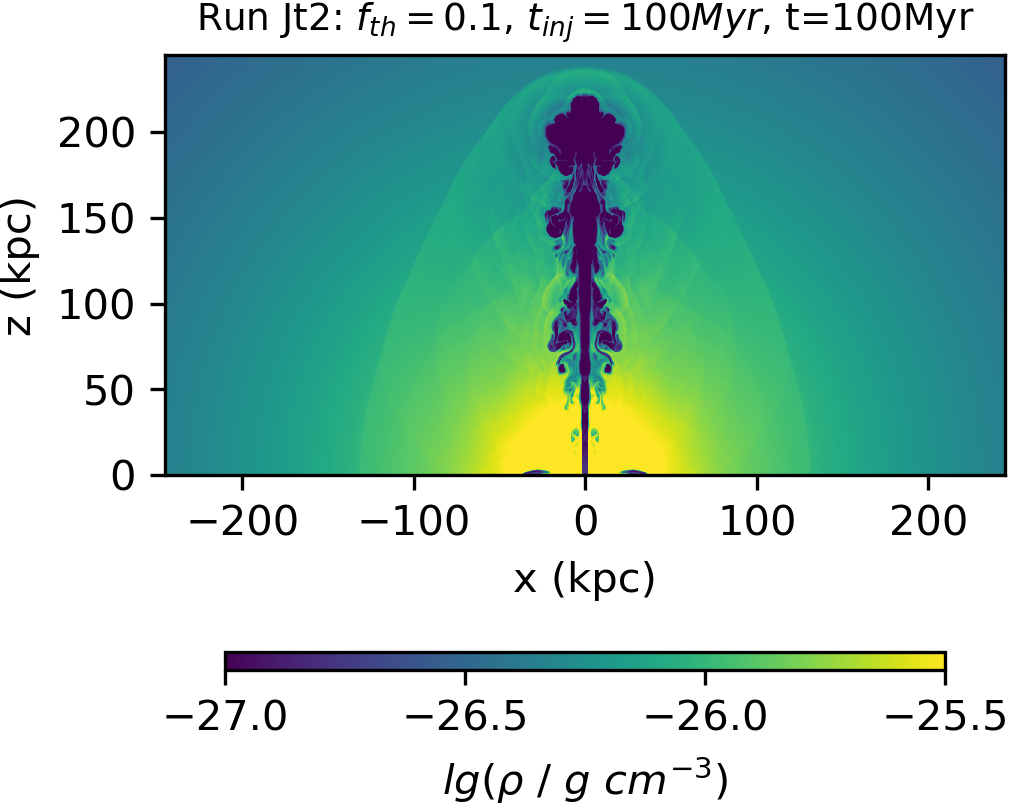}
}
\caption{The density distribution in run Jt2 at $t=t_{\rm inj}=100$ Myr when the jet injection just ends. The low-density ejecta bubble is in the braking phase. Weak compression fronts appearing like ripples can be clearly seen within the ejecta bubble.}
\label{plot:denJt2}
\end{figure*}

\begin{figure*}[h!]
\centering
\gridline{
\includegraphics[width=0.32\textwidth]{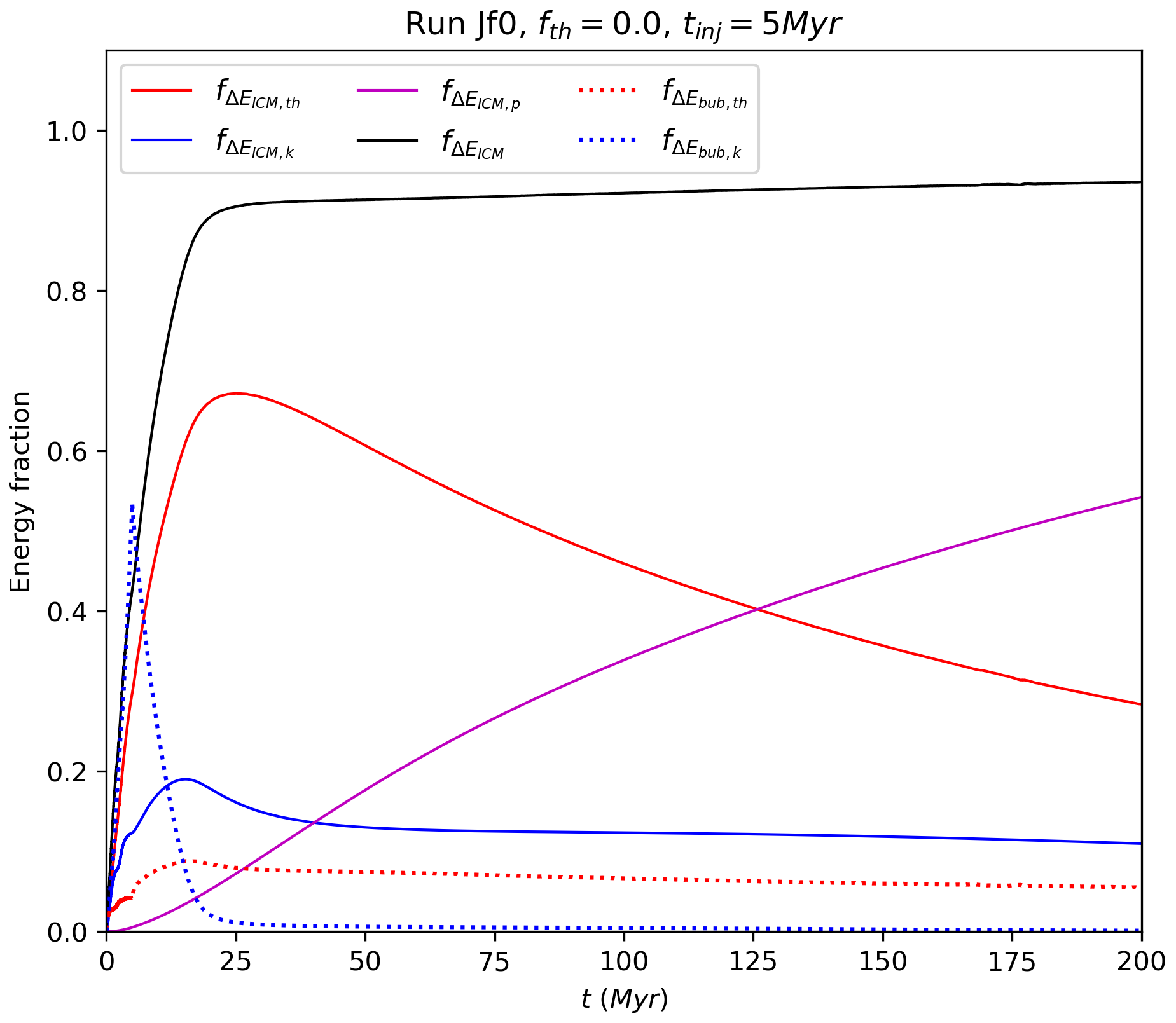}
\includegraphics[width=0.32\textwidth]{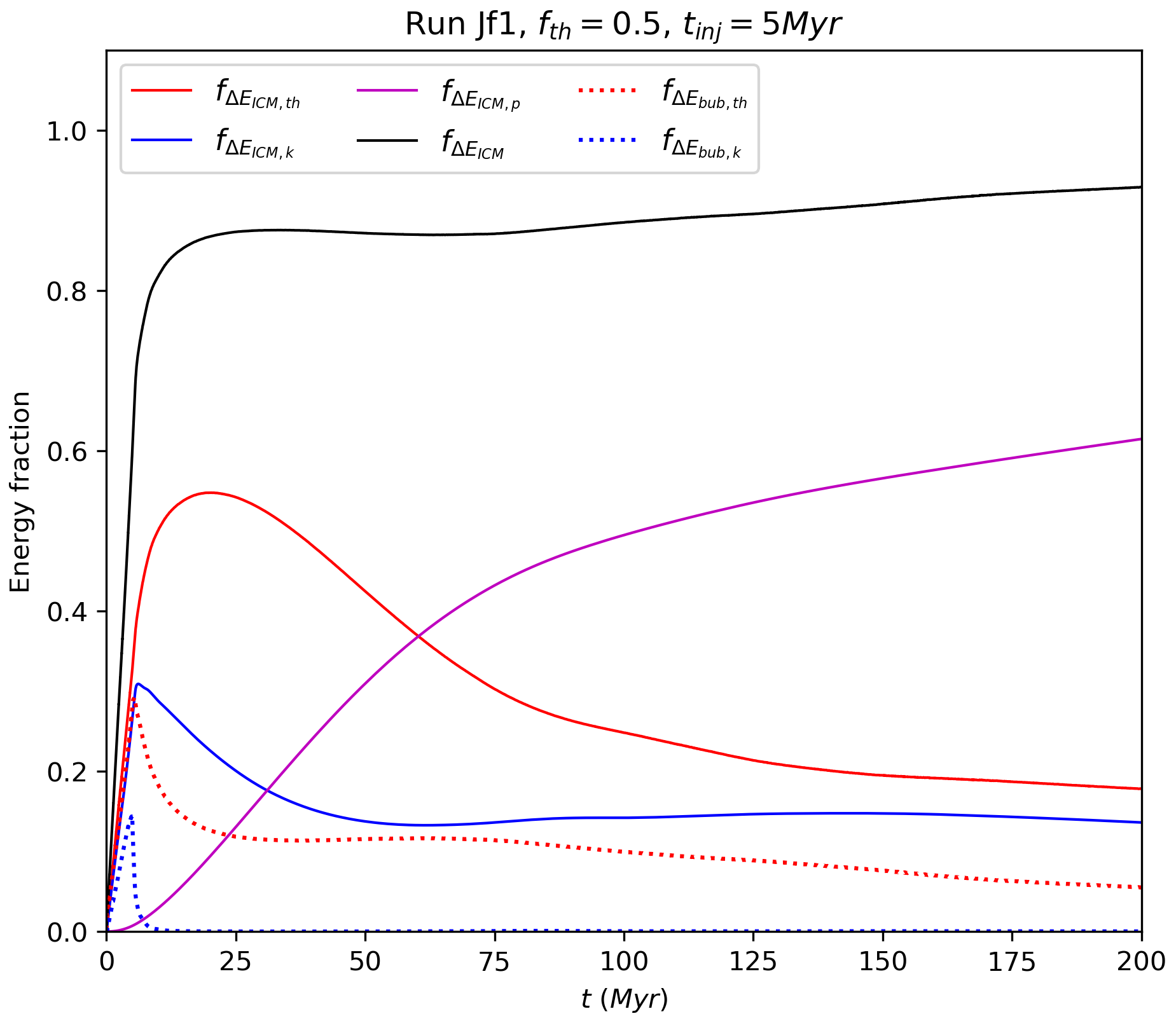}
\includegraphics[width=0.32\textwidth]{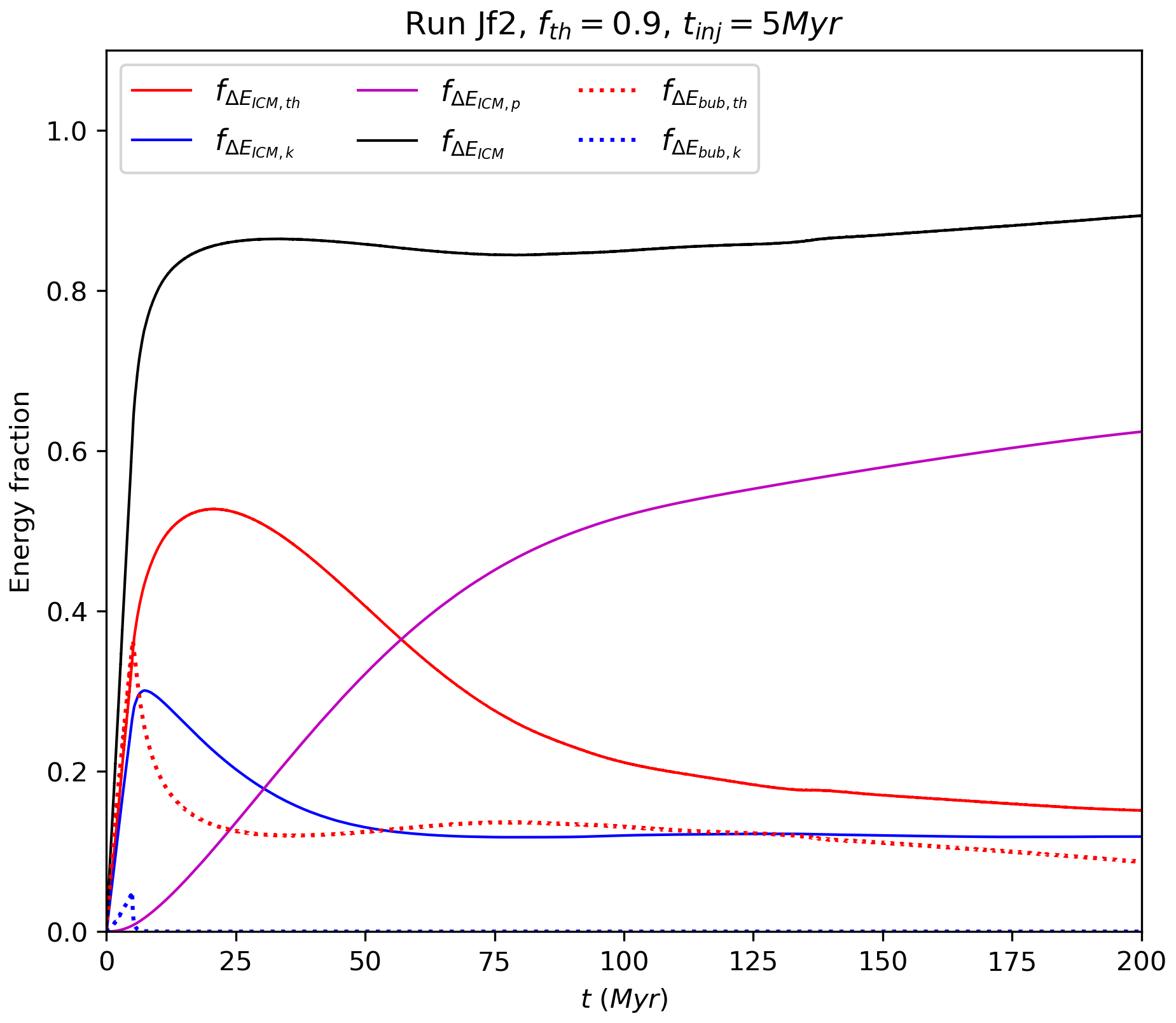}
}
\gridline{
\includegraphics[width=0.32\textwidth]{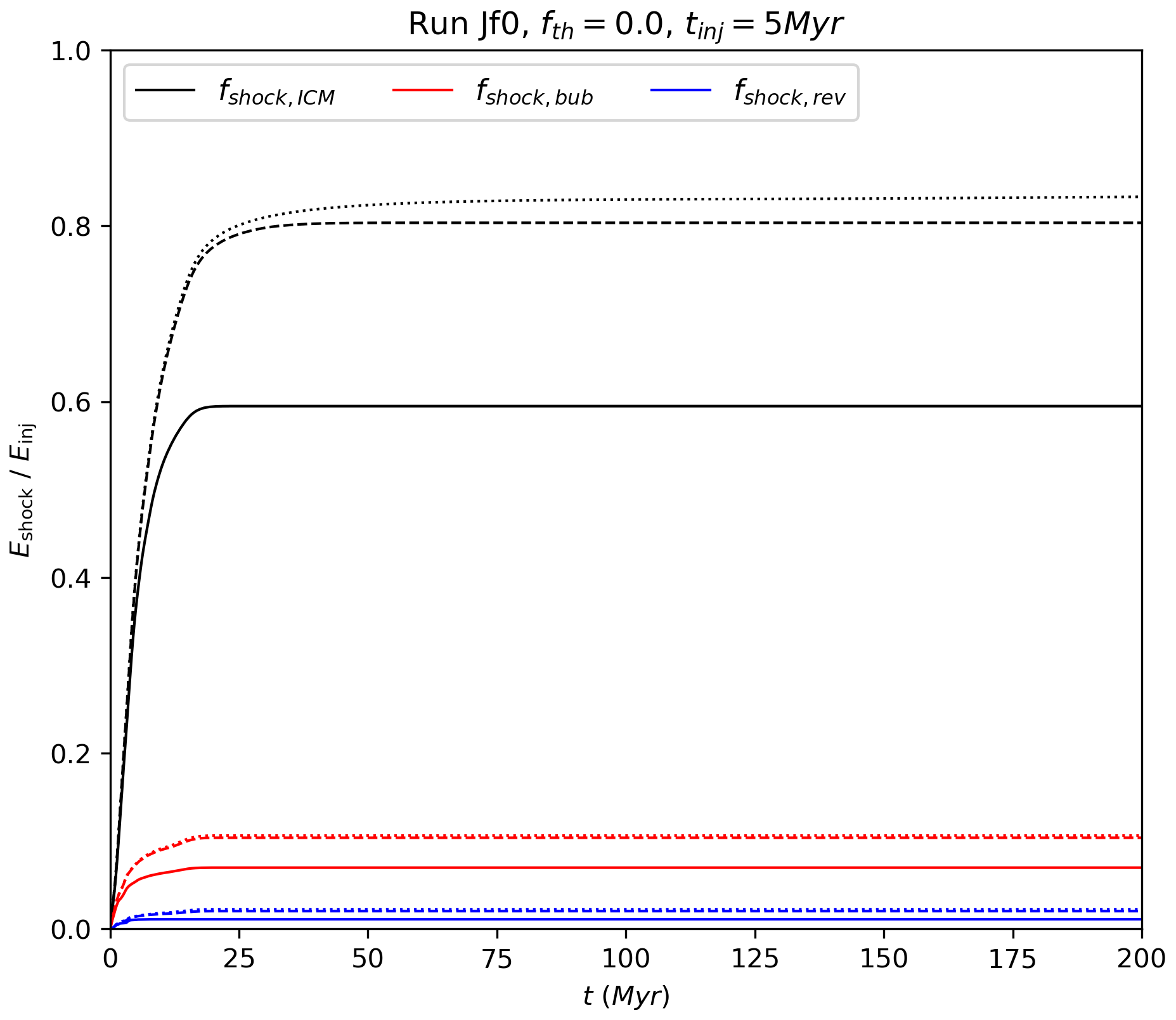}
\includegraphics[width=0.32\textwidth]{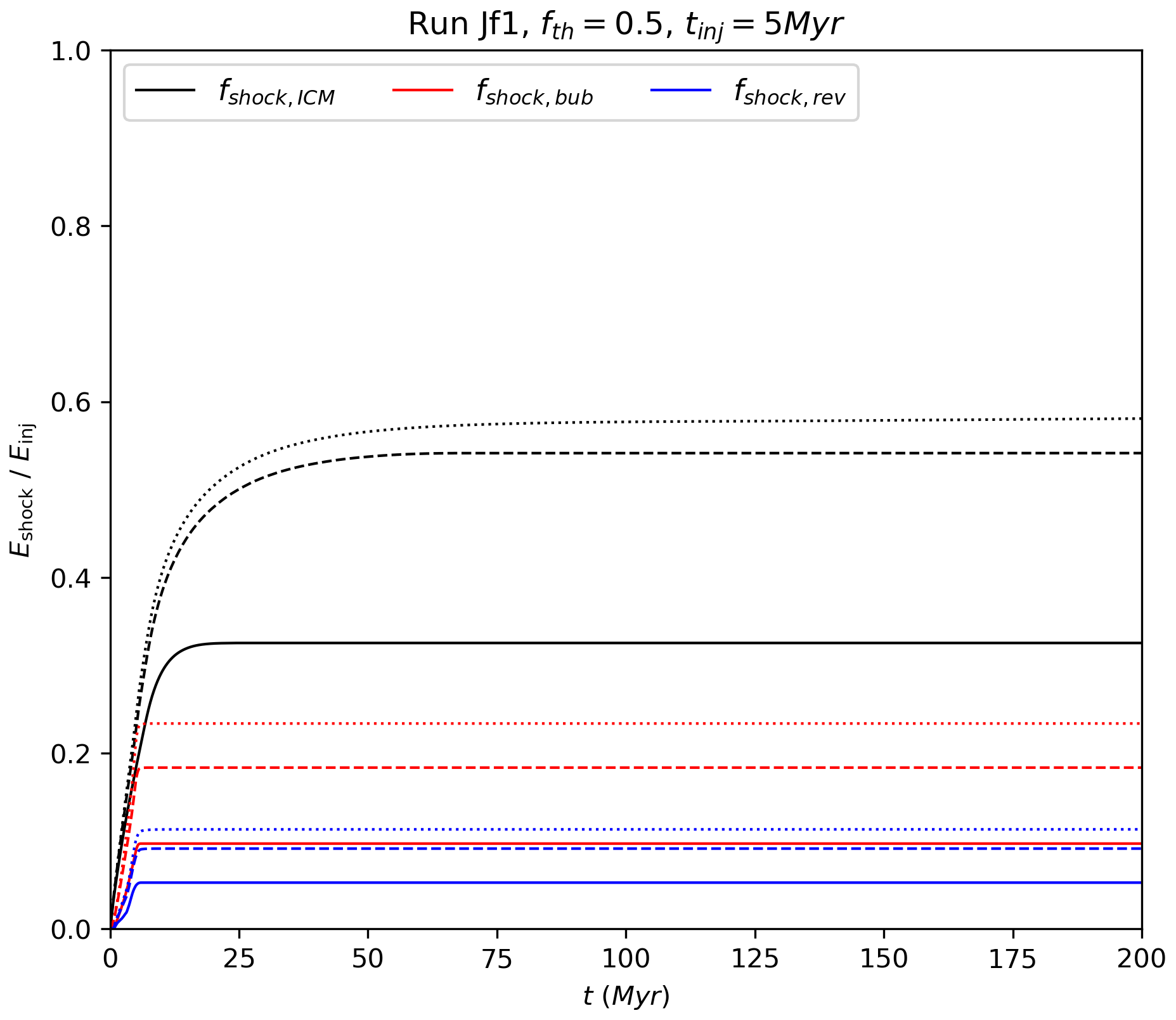}
\includegraphics[width=0.32\textwidth]{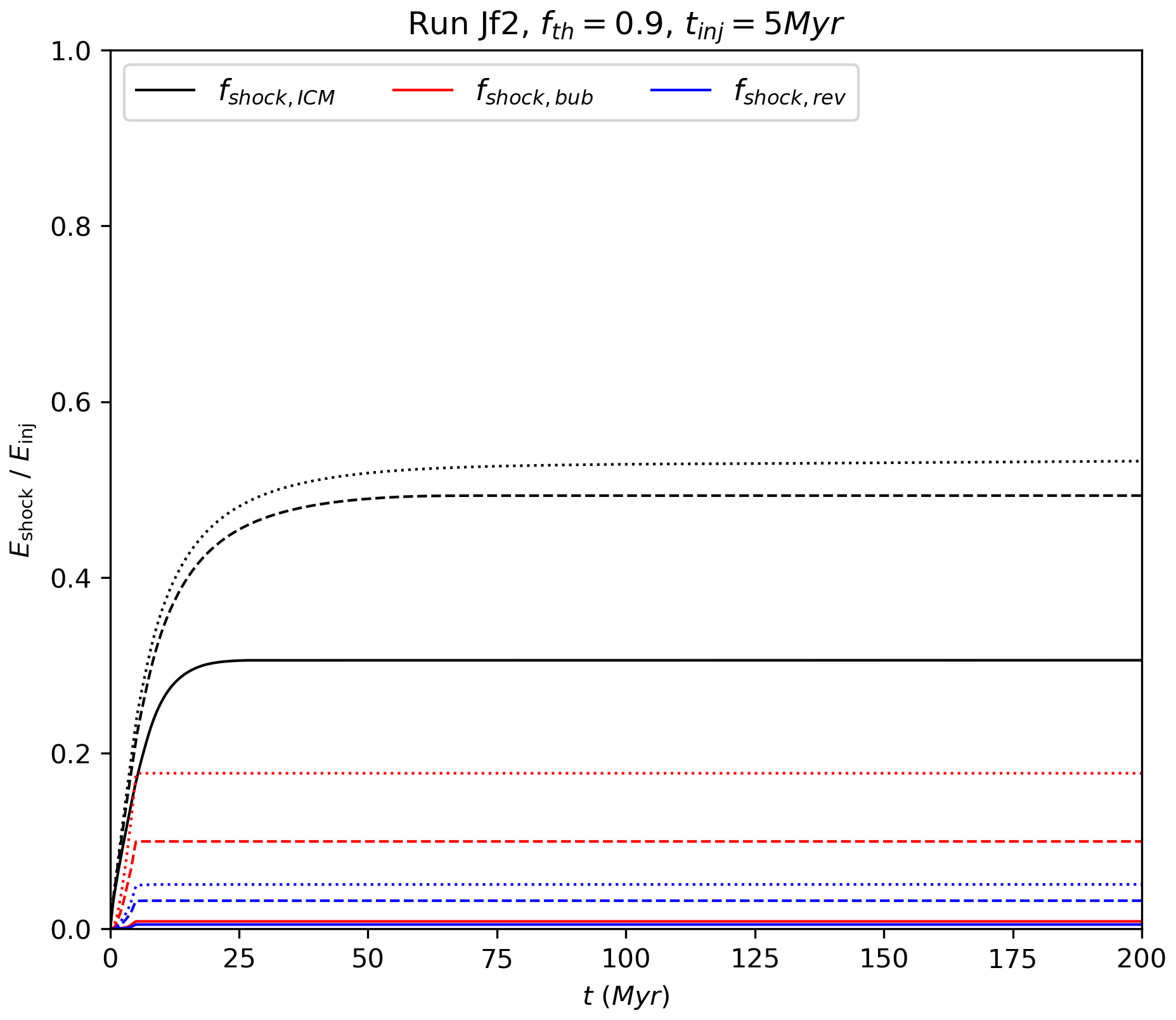}
}
\gridline{
\includegraphics[width=0.32\textwidth]{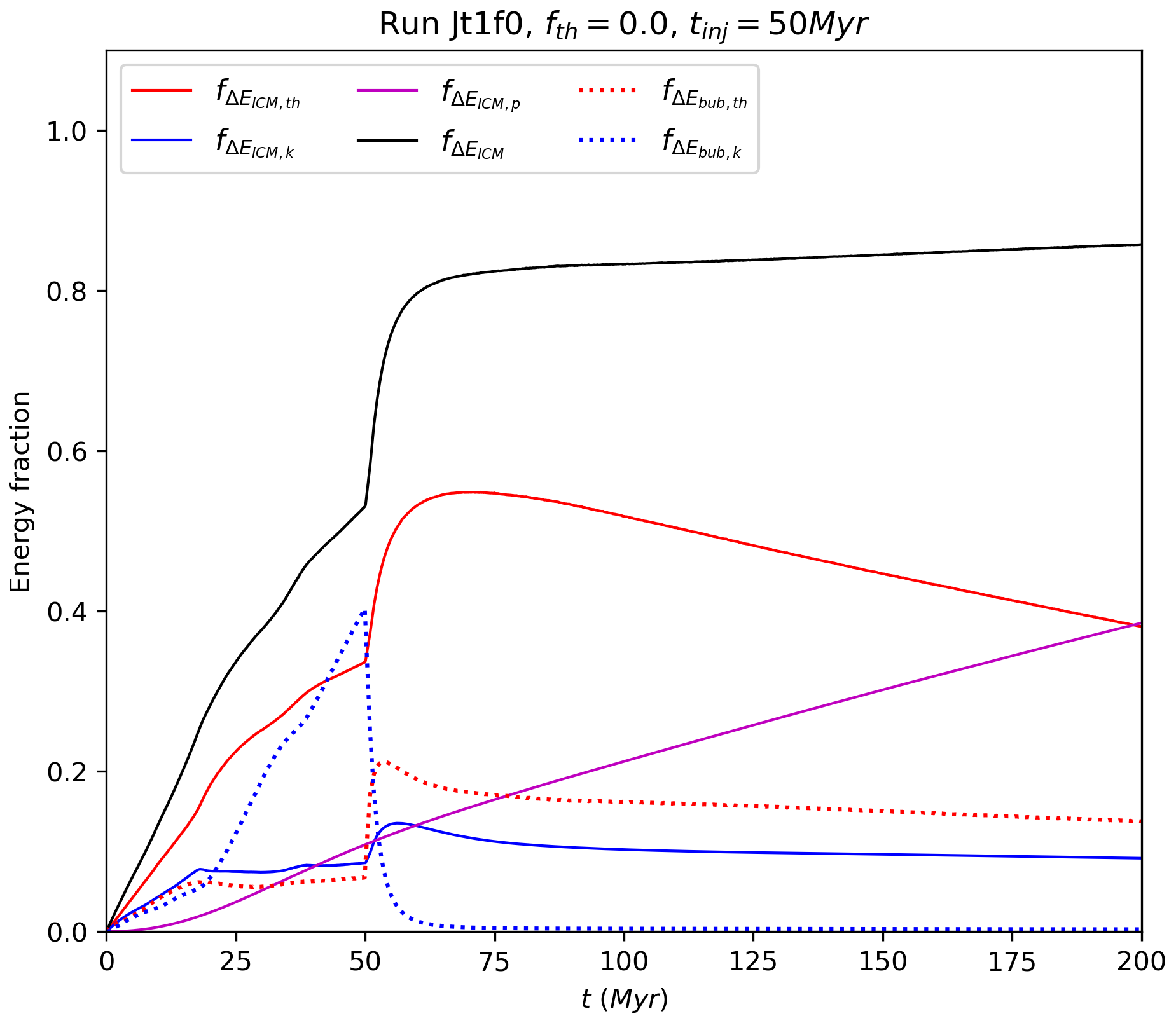}
\includegraphics[width=0.32\textwidth]{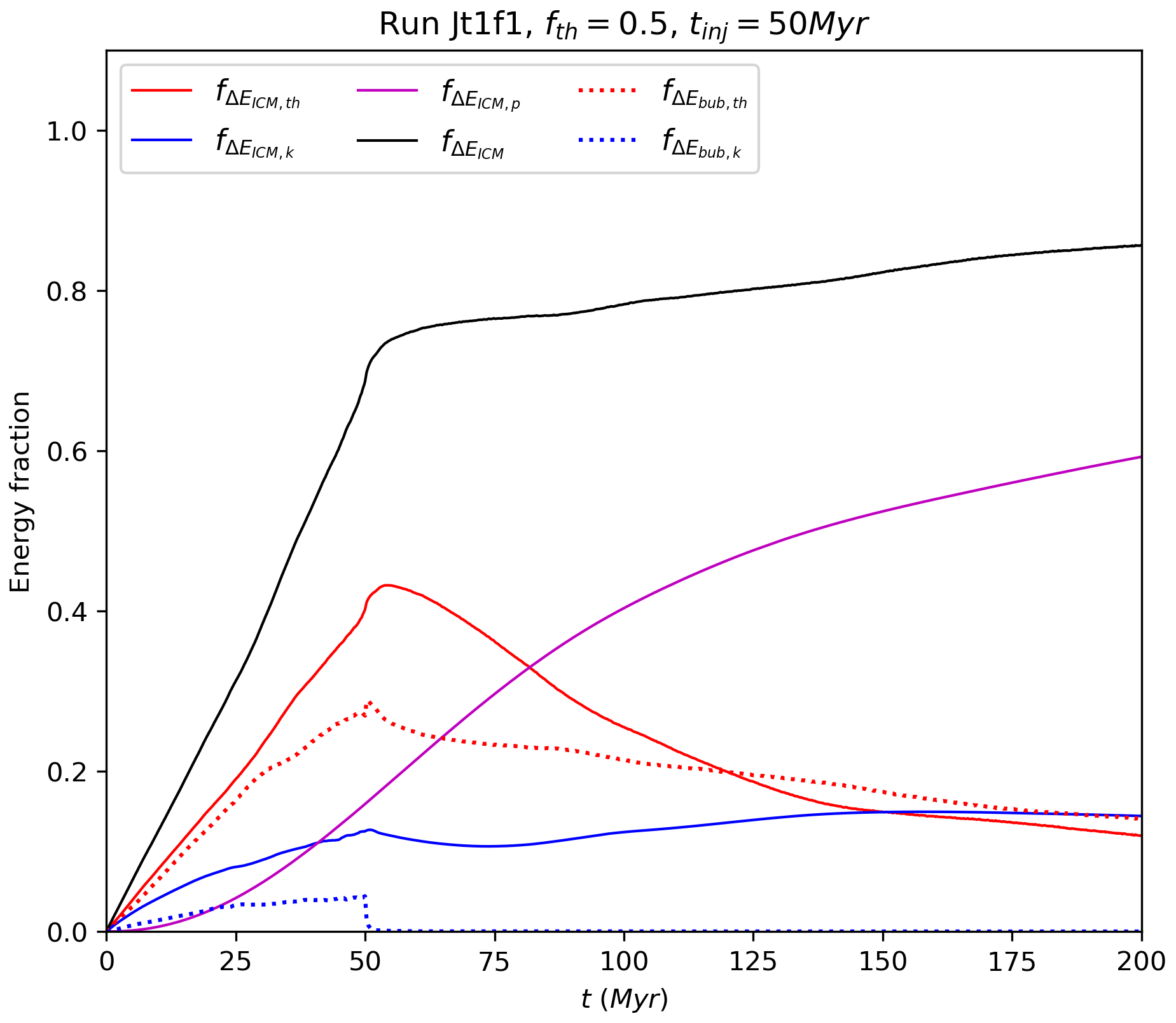}
\includegraphics[width=0.32\textwidth]{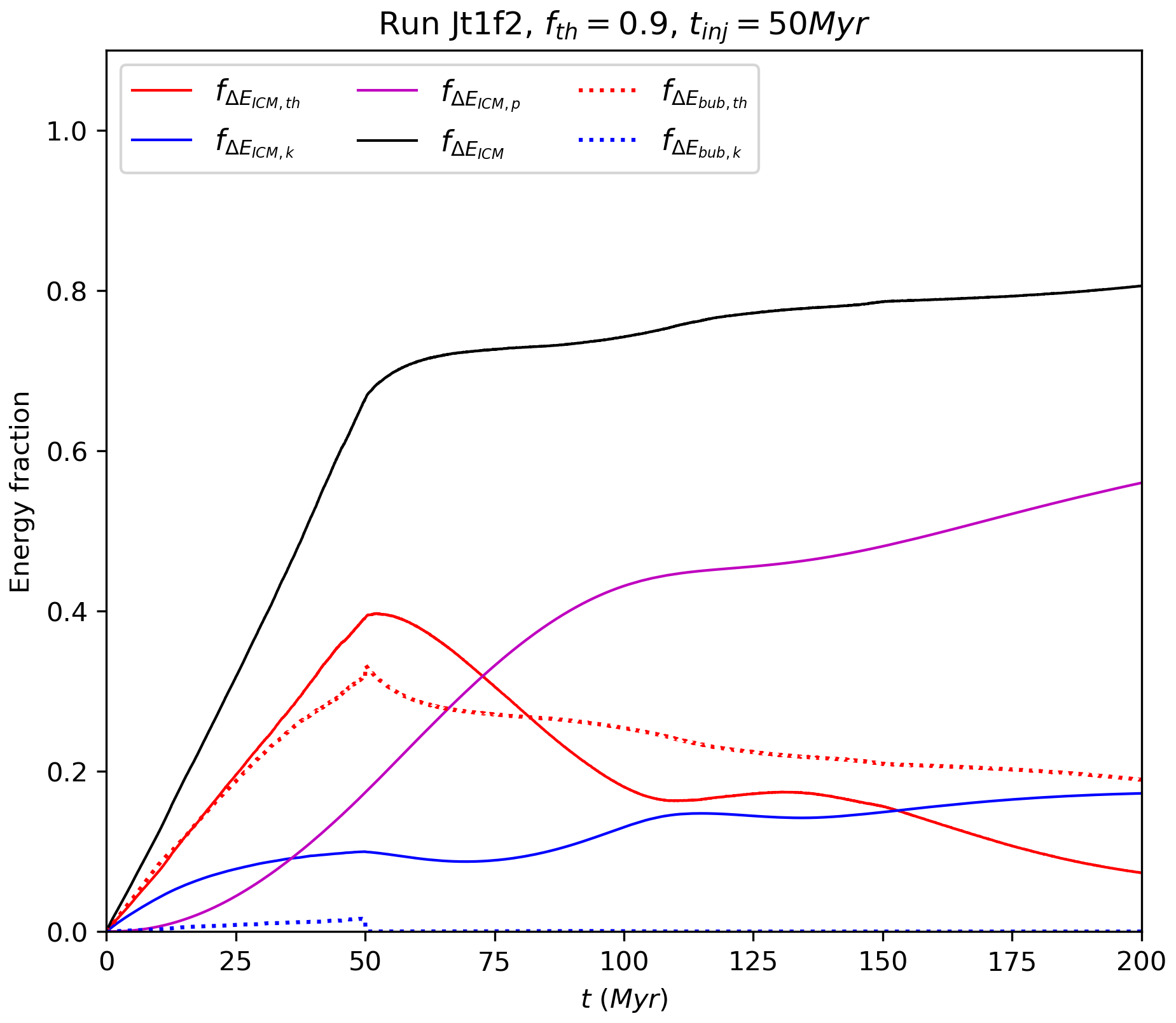}
}
\gridline{
\includegraphics[width=0.32\textwidth]{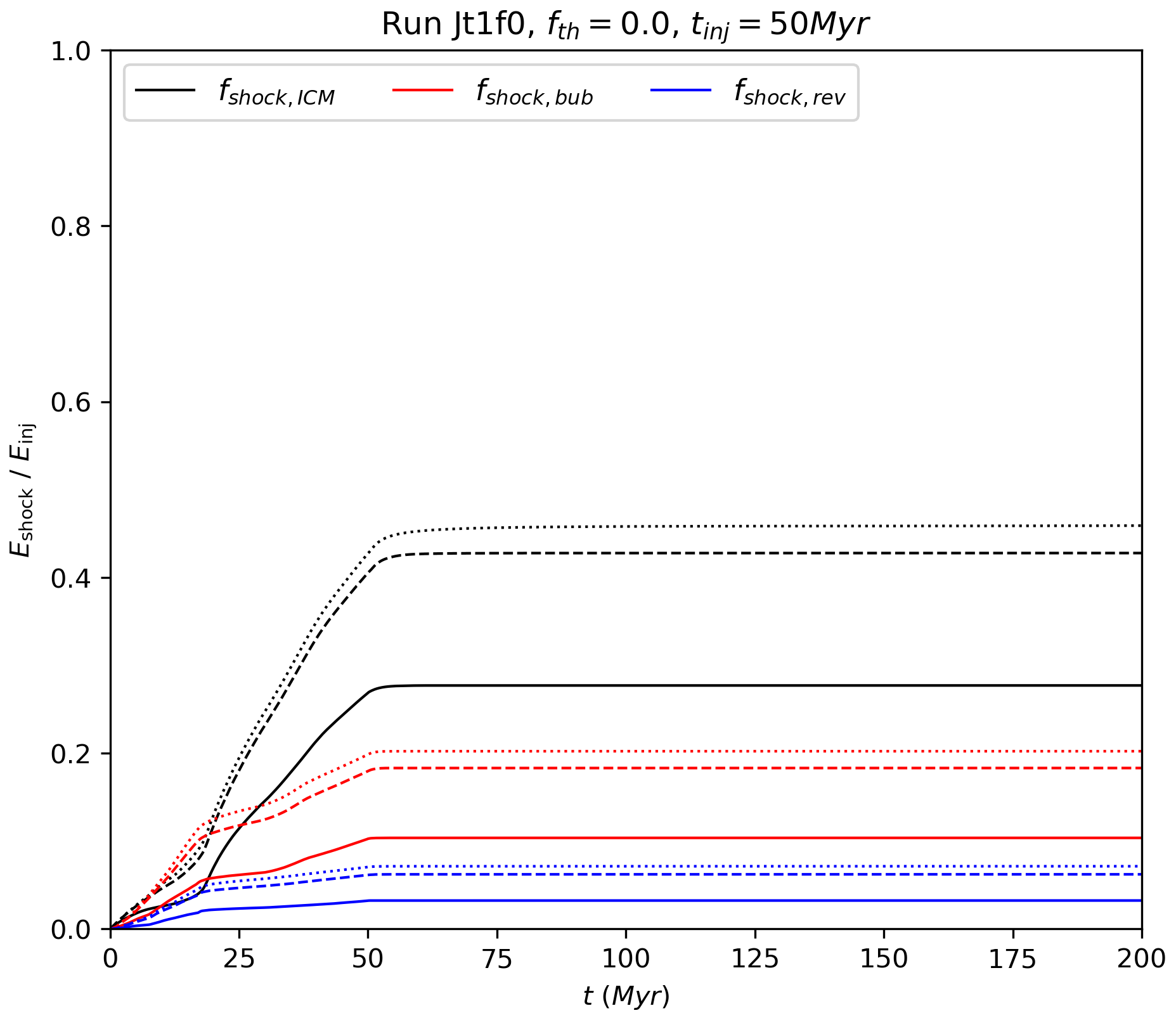}
\includegraphics[width=0.32\textwidth]{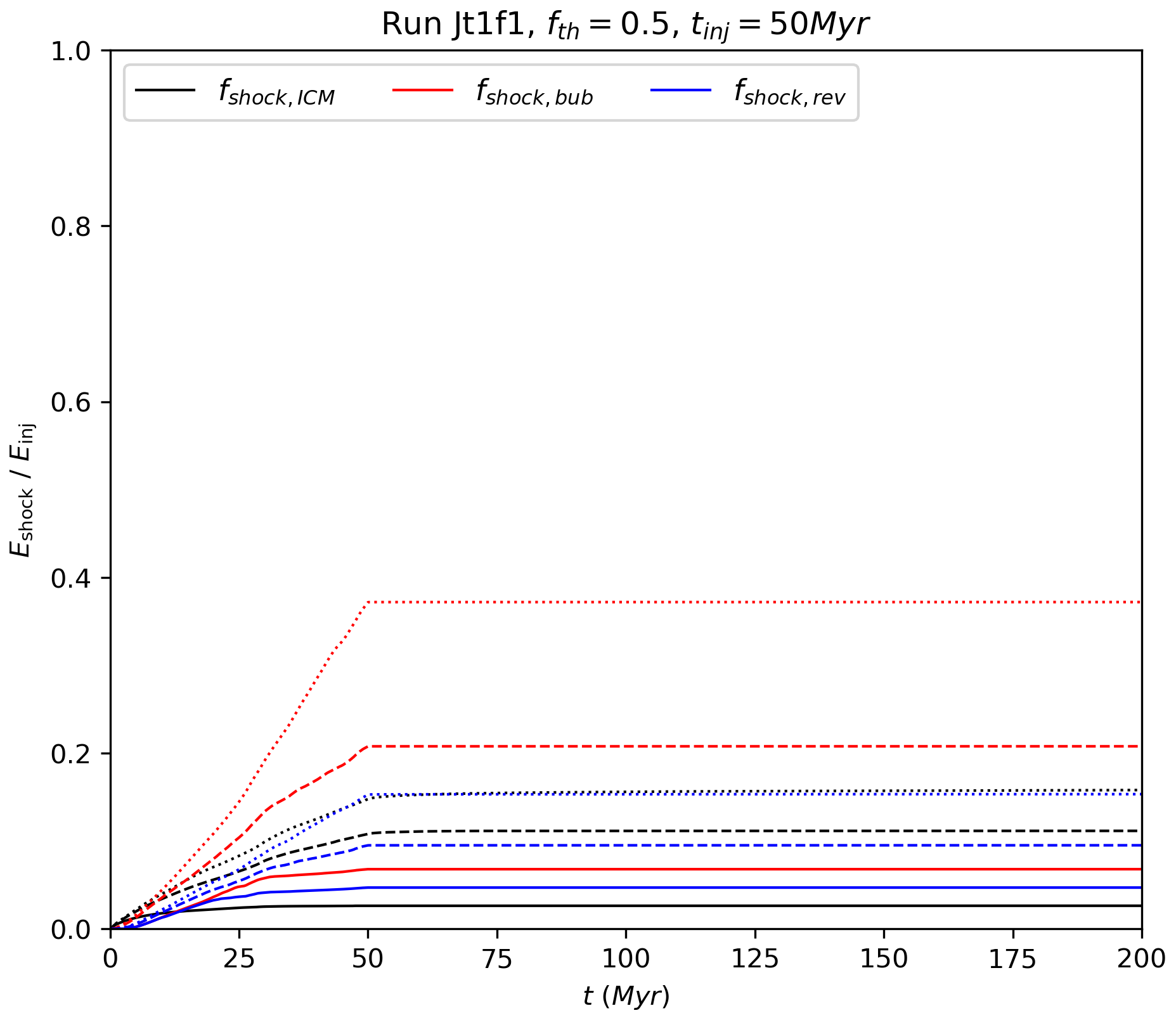}
\includegraphics[width=0.32\textwidth]{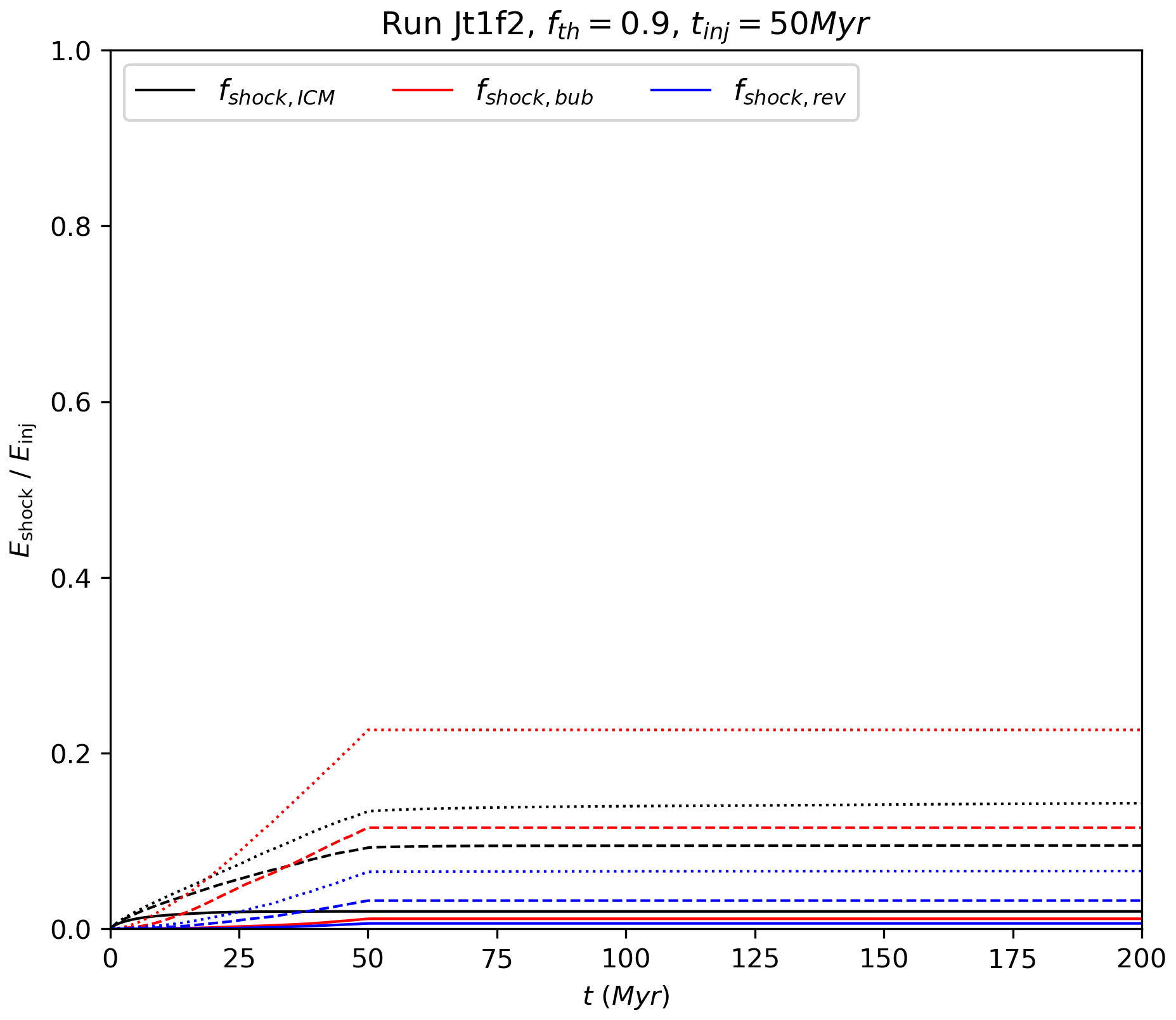}
}
\caption{Same as Figure 7, but showing the results of six additional runs. }
\label{plot:plot7more}
\end{figure*}

\begin{figure*}[h!]
\centering
\gridline{
\includegraphics[height=0.62\textheight]{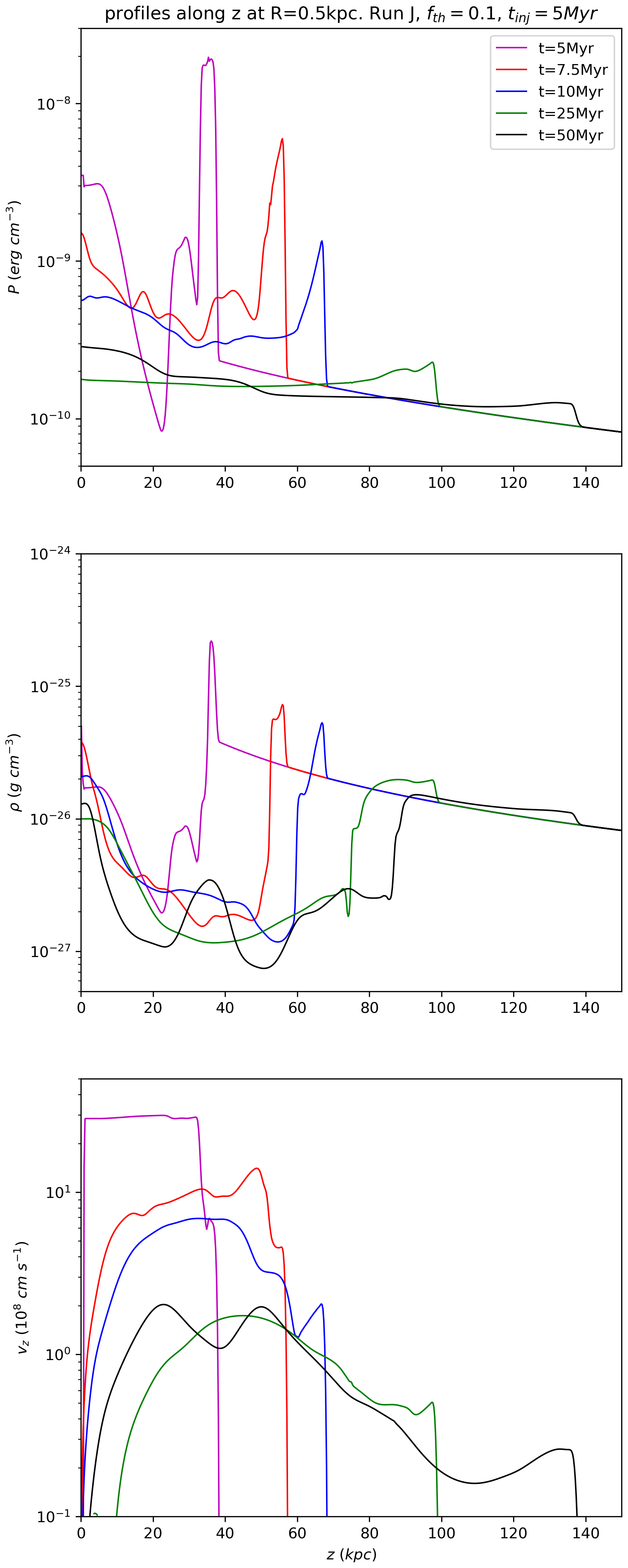}
\includegraphics[height=0.62\textheight]{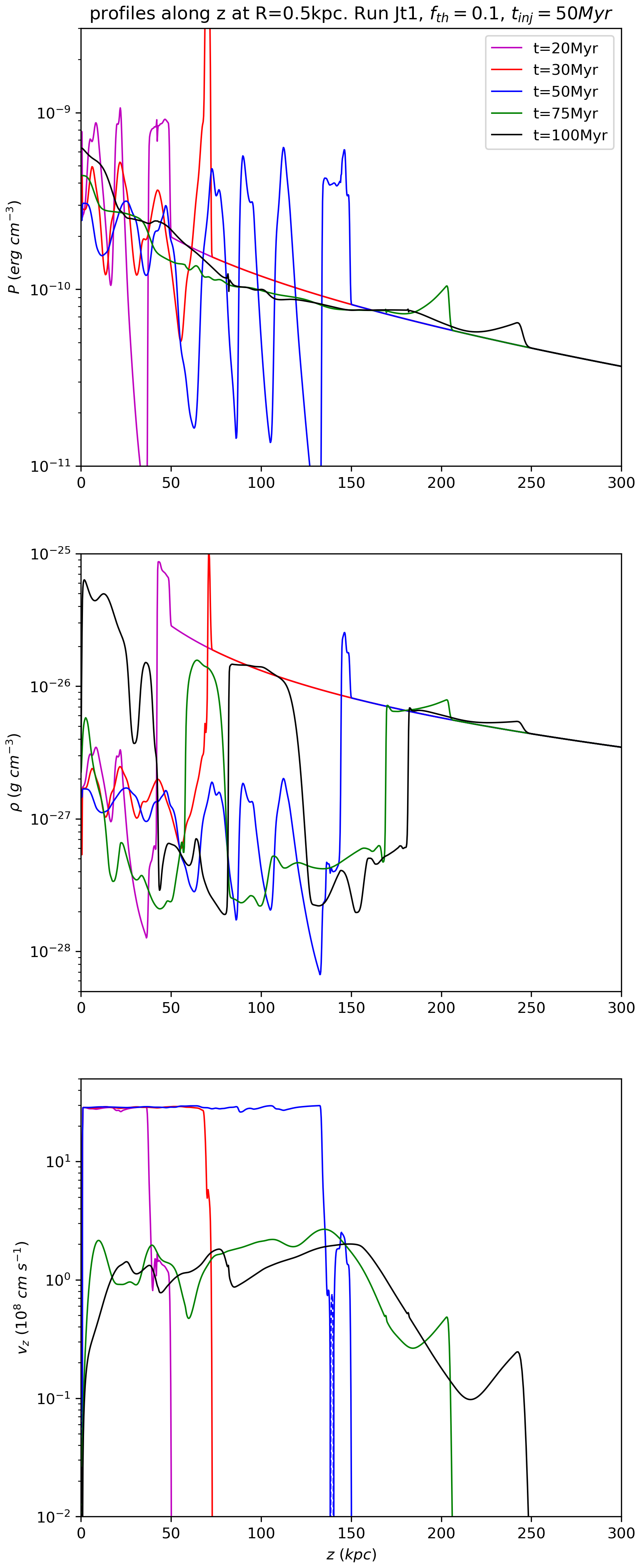}
\includegraphics[height=0.62\textheight]{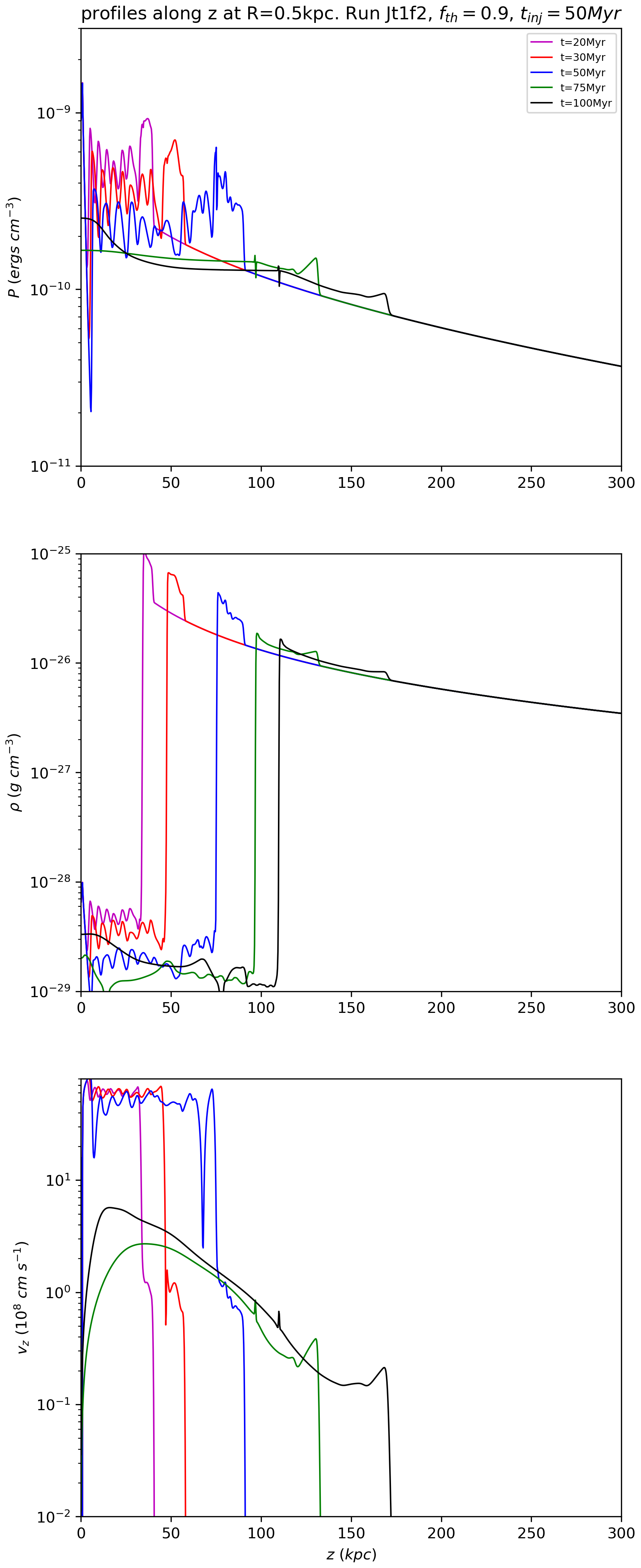}
}
\caption{Temporal evolution of profiles of gas pressure (top), density (middle) and the $z$-component velocity $v_{\rm z}$ (bottom) along the $z$ direction at $R=0.5$ kpc in three representative runs: run J (left), Jt1 (middle) and Jt1f2 (right). The dashed lines in the bottom panels refer to gas inflows with negative values of $v_{\rm z}$.}
\label{plot:shockx}
\end{figure*}

\begin{figure*}
\centering
\gridline{
\includegraphics[height=0.62\textheight]{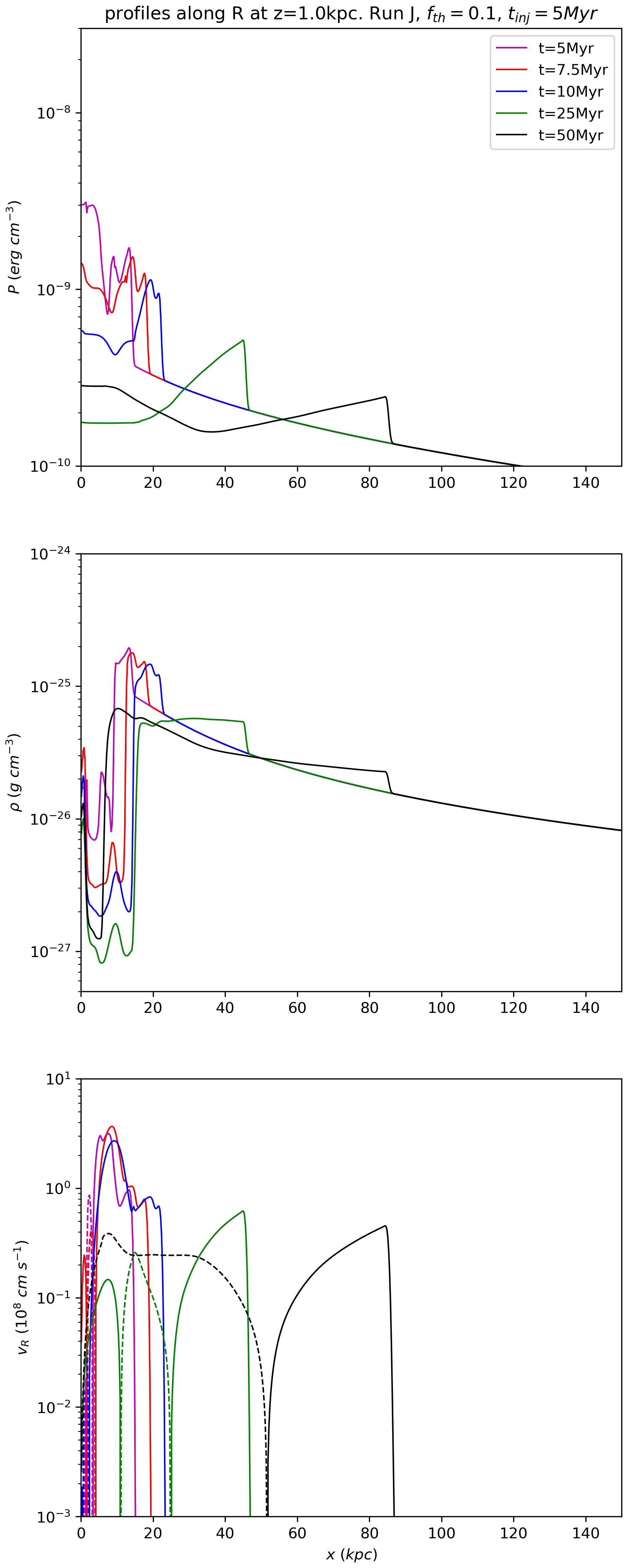}
\includegraphics[height=0.62\textheight]{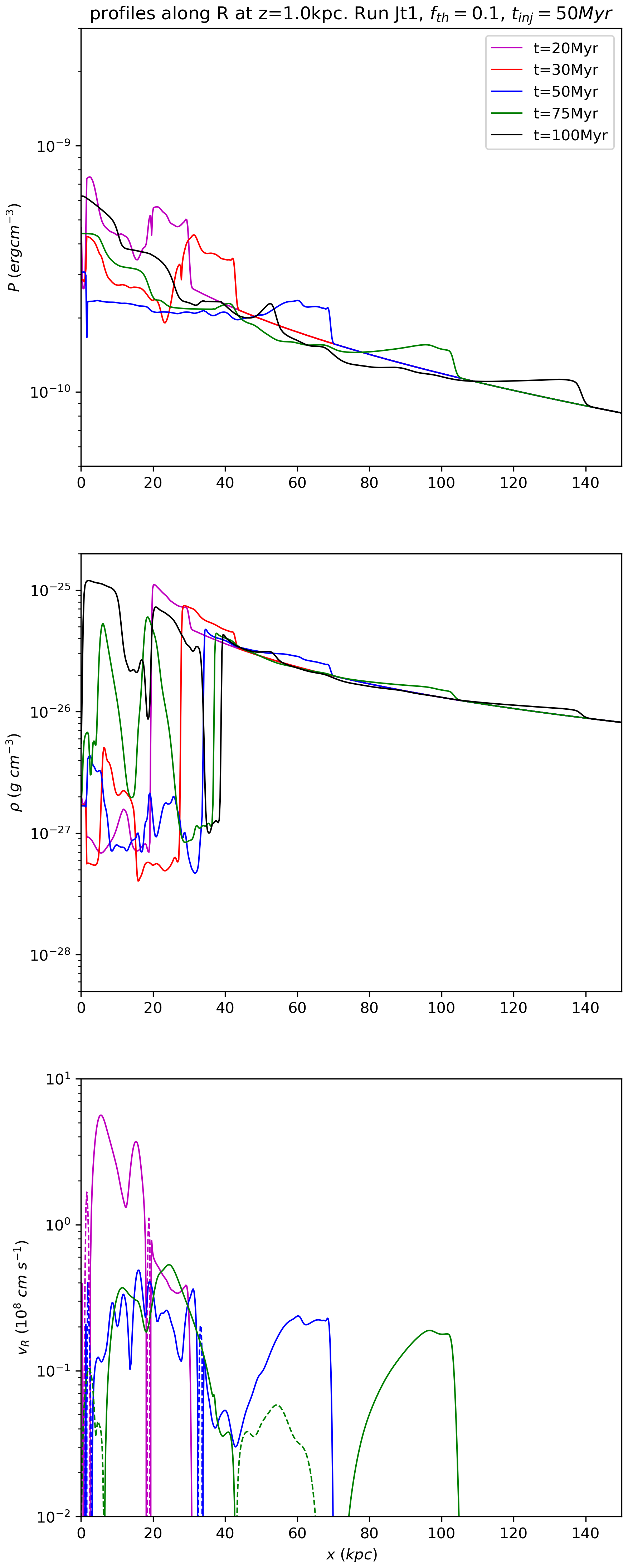}
\includegraphics[height=0.62\textheight]{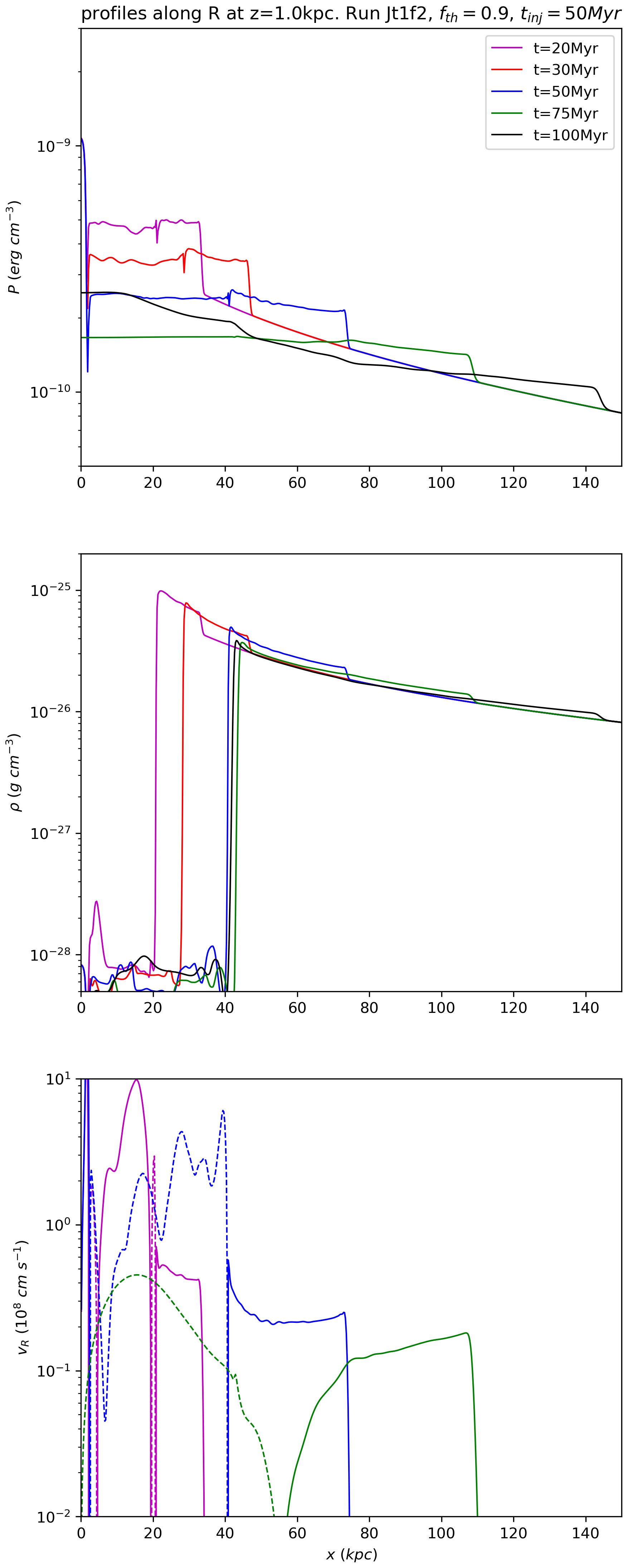}
}
\caption{Temporal evolution of profiles of gas pressure (top), density (middle) and the $R$-component velocity $v_{\rm R}$ (bottom) along the $R$ direction at $z=1$ kpc in three representative runs: run J (left), Jt1 (middle) and Jt1f2 (right). The dashed lines in the bottom panels refer to gas inflows with negative values of $v_{\rm R}$. In the bottom-middle and bottom-right panels, the lines corresponding to $t=30$ and $100$ Myr are omitted to avoid line crowding.}  
\label{plot:shocky}
\end{figure*}

\end{document}